\documentclass[10pt]{article} 

\usepackage[utf8]{inputenc}
\usepackage[T1]{fontenc}
\usepackage{times}

\usepackage{amsmath, amssymb, amsfonts, amsthm}

\usepackage[letterpaper,margin=3cm]{geometry}

\usepackage[pdftex]{graphicx}
\usepackage{float}
\usepackage{booktabs}
\usepackage{longtable}
\usepackage[tableposition=below]{caption}
\usepackage{subcaption}
\usepackage[export]{adjustbox}
\usepackage{rotating} 

\usepackage{natbib}

\usepackage[
  colorlinks=true,
  linkcolor=blue,
  urlcolor=blue,
  citecolor=blue
]{hyperref}

\makeatletter
\newcommand\setcurrentname[1]{\def\@currentlabelname{#1}}
\makeatother

\usepackage{algorithm}
\usepackage[noend]{algpseudocode}

\algdef{SE}[SUBALG]{Indent}{EndIndent}{}{\algorithmicend\ }%
\algtext*{Indent}
\algtext*{EndIndent}
\let\oldReturn\Return
\renewcommand{\Return}{\State\oldReturn}

\algnewcommand\algorithmicnot{\textbf{not}}
\algdef{SE}[IF]{IfNot}{EndIf}[1]{\algorithmicif\ \algorithmicnot\ #1\ \algorithmicthen}{\algorithmicend\ \algorithmicif}

\usepackage{titlesec}

\titleformat{\section}
  {\large\bfseries}
  {\thesection}{1em}{}
\titlespacing*{\section}{0pt}{1.25ex plus 0.5ex minus 0.2ex}{0.8ex}

\titleformat{\subsection}
  {\normalsize\bfseries}
  {\thesubsection}{1em}{}
\titlespacing*{\subsection}{0pt}{1.0ex plus 0.4ex minus 0.2ex}{0.6ex}

\titleformat{\subsubsection}
  {\normalsize\itshape}
  {\thesubsubsection}{1em}{}
\titlespacing*{\subsubsection}{0pt}{0.8ex plus 0.3ex minus 0.2ex}{0.5ex}

\renewcommand{\vec}[1]{\ensuremath{\boldsymbol{#1}}}

\makeatletter
\renewcommand{\maketitle}{%
  \begin{center}
    {\Large\bfseries \@title \par}
    \vskip 0.9em
    {\normalsize \@author \par}
  \end{center}
  \vskip 1.0em
}
\makeatother

\date{}

\title{Physics-constrained symbolic regression for discovering closed-form equations of multimodal water retention curves from experimental data}

\author{%
Yejin Kim$^{1}$ \quad Hyoung Suk Suh$^{1,*}$\\[0.6ex]
$^{1}$Department of Civil and Environmental Engineering, Case Western Reserve University, United States\\[0.4ex]
\texttt{y.kim@case.edu} \quad \texttt{hssuh@case.edu}
}

\begin{document}

\maketitle

\begingroup
\renewcommand\thefootnote{*}
\footnotetext{Corresponding author.}
\endgroup

\begin{abstract}
Modeling the unsaturated behavior of porous materials with multimodal pore size distributions presents significant challenges, as standard hydraulic models often fail to capture their complex, multi-scale characteristics.
A common workaround involves superposing unimodal retention functions, each tailored to a specific pore size range; however, this approach requires separate parameter identification for each mode, which limits interpretability and generalizability, especially in data-sparse scenarios.
In this work, we introduce a fundamentally different approach: a physics-constrained machine learning framework designed for meta-modeling, enabling the automatic discovery of closed-form mathematical expressions for multimodal water retention curves directly from experimental data.
Mathematical expressions are represented as binary trees and evolved via genetic programming, while physical constraints are embedded into the loss function to guide the symbolic regressor toward solutions that are physically consistent and mathematically robust.
Our results demonstrate that the proposed framework can discover closed-form equations that effectively represent the water retention characteristics of porous materials with varying pore structures.
To support third-party validation, application, and extension, we make the full implementation publicly available in an open-source repository.
\end{abstract}

\section{Introduction}
\label{intro}
Water retention characteristics play a central role in modeling unsaturated flow in porous media, as they govern how fluids are distributed and retained under varying conditions. 
This behavior can be captured by the water retention curve, which links matric suction to degree of saturation, and serves as a key constitutive equation in macroscopic models for hydro-mechanically coupled simulations \citep{borja2004cam, griffiths2005unsaturated, song2014mathematical, kim2025can}. 
Semi-empirical models, such as those proposed by \citep{brooks1965hydraulic, van1980closed, kosugi1994three, fredlund1994equations}, are commonly used in these simulations, as they provide simple analytical expressions with relatively few parameters to describe the water retention curve. 
However, these models rely on the assumption that the corresponding porous medium exhibits a unimodal pore size distribution, making them unsuitable for accurately capturing unsaturated behavior of materials with complex or multimodal pore structures \citep{li2014predicting, zhou2017bimodal}. 
A typical workaround is to combine multiple unimodal models, each corresponding to a distinct pore size range \citep{durner1994hydraulic, priesack2006closed}, but this approach may complicate the parameter identification process and thus limits its applicability, particularly when the available data are sparse.

Water retention characteristics play a central role in modeling unsaturated flow in porous media, as they govern how fluids are distributed and retained under varying conditions. 
This behavior can be captured by the water retention curve, which links matric suction to degree of saturation, and serves as a key constitutive equation in macroscopic models for hydro-mechanically coupled simulations \citep{borja2004cam, griffiths2005unsaturated, song2014mathematical, kim2025can}. 
Semi-empirical models, such as those proposed by \citep{brooks1965hydraulic, van1980closed, kosugi1994three, fredlund1994equations}, are commonly used in these simulations, as they provide simple analytical expressions with relatively few parameters to describe the water retention curve. 
However, these models rely on the assumption that the corresponding porous medium exhibits a unimodal pore size distribution, making them unsuitable for accurately capturing unsaturated behavior of materials with complex or multimodal pore structures \citep{li2014predicting, zhou2017bimodal}. 
A typical workaround is to combine multiple unimodal models, each corresponding to a distinct pore size range \citep{durner1994hydraulic, priesack2006closed}, but this approach may complicate the parameter identification process and thus limits its applicability, particularly when the available data are sparse.

One possible direction to overcome the aforementioned challenges is to adopt data-driven methods that can learn complex behaviors directly from experimental data without relying on pre-specified functional forms or the explicit calibration of material parameters. 
Among various approaches, deep neural networks have emerged as a popular choice due to their high flexibility and expressive capability. 
They have demonstrated strong potential to reproduce a wide range of constitutive models, for instance, elastic energy functional \citep{vlassis2020geometric, linden2023neural}, plastic yield surfaces and hardening rules \citep{vlassis2021sobolev, suh2023publicly}, permeabilities \citep{singh2020modelling, tembely2020deep}, and also water retention characteristics \citep{liu2019application, heider2021offline}. 
Despite their powerful capacity to represent such complex, nonlinear relationships, employing neural networks has not yet become mainstream in practical engineering applications. 
This is largely due to their lack of interpretability; the black-box nature of neural networks often makes the trustworthiness of the trained model questionable, which may have hindered their broader use \citep{murdoch2019definitions, fan2020interpretability, wing2021trustworthy}.

Although several efforts have been made to improve the interpretability of machine learning approaches \citep{saleem2022explaining, li2022interpretable, liu2024kan}, as noted by \citep{agarwal2021neural}, models designed for improved interpretability often suffer from a lack of expressiveness needed to achieve high predictive accuracy (i.e., interpretability-accuracy trade-off). 
In contrast, symbolic regression--which aims to discover mathematical expressions that best fit a given dataset--can achieve both high interpretability and accuracy, provided the combinatorial optimization used to search the expression space is successful \citep{angelis2023artificial, cranmer2023interpretable, makke2024interpretable}. 
This best-case outcome, however, is often difficult to achieve in practice. 
One major limitation of symbolic regression is its scalability: as the dimensionality of the input space or the complexity of the target function increases, the number of candidate expressions grow exponentially, making the search computationally demanding \citep{petersen2019deep, udrescu2020ai}. 
Moreover, symbolic regression is particularly sensitive to noise and outliers in the data, which can lead to overfitting and the discovery of expressions that fail to generalize beyond the training dataset \citep{bomarito2022bayesian, de2023reducing}. 
To resolve these issues, recent studies \citep{bahmani2024discovering, suh2024data} have proposed hybrid frameworks that integrate neural networks with symbolic regression, wherein a neural network is first trained to approximate the data, followed by symbolic regression to extract a closed-form expression for the neural network function. 
While this approach has made partial success in resolving these issues, one thing that has been overlooked is that the resulting mathematical expression or its neural network counterpart may yield a learned function that is not physically sound (e.g., a learned water retention model that predicts saturation values exceeding unity or exhibits non-monotonic behavior).

Therefore, this work presents a meta-modeling approach for water retention in porous media using a multi-objective symbolic regressor that can discover closed-form mathematical expressions for multimodal water retention curves from experimental data, while maintaining physical consistency. 
The framework represents mathematical expressions as binary trees and employs genetic programming to evolve them over successive generations. 
A multi-objective optimization strategy drives this evolution by minimizing the prediction error while simultaneously enforcing physical constraints through dedicated loss terms, including monotonicity, limiting, and mode constraints. 
This approach enables the data-driven discovery of mathematical expressions that best describe the observed data while rigorously satisfying imposed physics constraints. 
Even when the resulting equations exhibit a certain degree of complexity, they remain transparent, interpretable, and analytically tractable, standing in clear contrast to black-box neural network models.
Consequently, the discovered closure relationships can be seamlessly integrated into existing hydraulic simulation codes in the same manner as conventional water retention models. 
To promote transparency, reproducibility, and further development, the full implementation is publicly available in an open-source repository. 
For clarity and focus, this study specifically addresses the modeling of water retention behavior during primary drying (i.e., drainage), deliberately excluding hysteresis effects, with the intention of addressing these complexities in future work.

\section{Method}
\label{sec:method}
This section begins by formulating the problem for discovering water retention models from experimental data as a multi-objective regression task in Section \ref{sec:learn_prob}. 
We then outline the key physical constraints--such as monotonicity, limiting conditions at extreme suction pressures, and saturation bounds--that the model must satisfy to accurately reflect the behavior of unsaturated porous media (Section \ref{sec:swrc_physics}). 
Finally, in Section \ref{sec:pcsr}, we present a physics-constrained symbolic regression (PCSR) framework for multimodal water retention curves, with physics constraints enforced during training via penalization. 
The performance of the proposed framework will be demonstrated through a series of examples later in Section \ref{sec:results}.

\subsection{Learning problem}
\label{sec:learn_prob}
The machine learning task considered in this work is to discover a function $\hat{S}_w(s)$ that maps a suction pressure $s \in \mathbb{R}$ to the corresponding water saturation ratio $S_w \in \mathbb{R}$, given a set of data pairs $\lbrace (s^{(i)}, S_w^{(i)}) \rbrace_{i=1}^{n_\text{data}}$, with $n_\text{data}$ representing the number of experimental measurements. 
Our goal is to learn a model that not only fits the observed data but also generalizes well to unseen matric suction levels, accurately representing the water retention behavior of porous materials. 
This requires ensuring that the learned function is consistent with the underlying physics. 
Furthermore, we assume that the shape of the target function is known a priori and characterized by a designated number of modes, denoted as $N_\text{mode}$. 
This parameter serves as a structural prior, identifiable from experimental observations or material characteristics, and guides the symbolic search to mitigate overfitting without prescribing a rigid mathematical architecture.

To achieve this, one may frame the problem as a multi-objective optimization task, in which the total loss function $\mathcal{L}$ comprises three major components: a data loss term $\mathcal{L}_{\text{data}}$ to ensure the fitness to the experimental data, a physics loss term $\mathcal{L}_{\text{phys}}$ to enforce essential physics constraints, and a shape-guiding loss term $\mathcal{L}_{\text{mode}}$ to achieve the desired number of modes in the learned function: 
\begin{equation}
\label{eq:loss_func}
\mathcal{L} = 
\underbrace{\frac{1}{n_\text{data}} \sum_{i=1}^{n_\text{data}} \left[ \hat{S}^{(i)}_w(s^{(i)}) - S^{(i)}_w \right]^2}_{= \mathcal{L}_\text{data}} + \mathcal{L}_\text{phys} + \mathcal{L}_\text{mode},
\end{equation}
where the quantities superposed with a hat correspond to predictions of the learned model, while $N_\text{mode}$ represents the target number of modes for the learned function. 
If neural networks are employed to represent the model, for instance, all relevant $\mathcal{L}_\text{phys}$ terms must be formulated as (partially) differentiable functions in order to integrate them into the backpropagation process during optimization. 
In contrast, the symbolic regression approach adopted in this work does not require the loss components to be differentiable. 
As such, it is not necessary to define a specific functional form for $\mathcal{L}_\text{phys}$; instead, the relevant physics constraints are incorporated through a custom penalization strategy.  
The constraints themselves will be summarized in Section~\ref{sec:swrc_physics}, and implementation details will be presented in Section~\ref{sec:pcsr}.

\begin{figure}[h]
\centering
\noindent\includegraphics[width=0.45\textwidth]{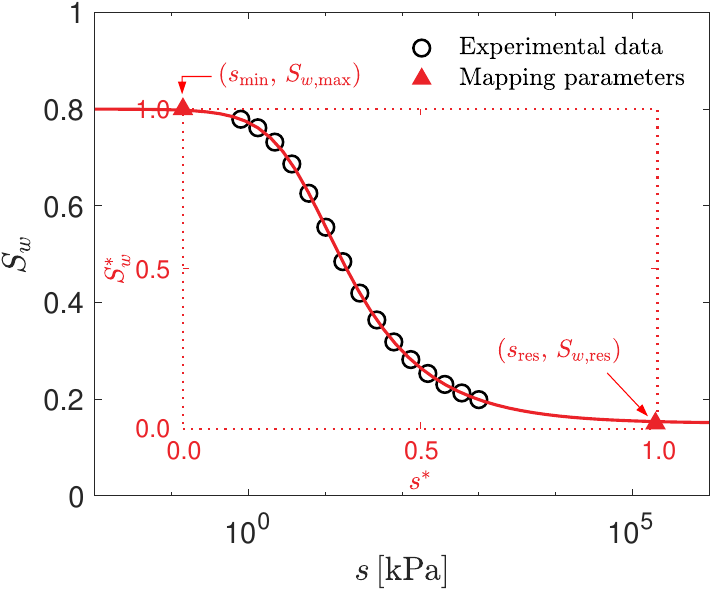}
\caption{Schematic illustration of the data mapping procedure. Black hollow symbols represent exemplary water retention data, with the black axes denoting the original $(s, S_w)$ space and the red axes indicating the mapped $(s^*, S^*_w)$ space. The upper left corner of the red box corresponds to the reference point $(s_\text{min}, S_{w,\text{max}})$ and the lower right corner to $(s_\text{res}, S_{w,\text{min}})$ in the original space, which map to $(0,1)$ and $(1,0)$, respectively, in the transformed space. This mapping ensures that the learned function (red curve) resides within $s^* \in [0,1]$ and $S^*_w \in [0, 1]$.}
\label{fig:training_space}
\end{figure}

While the function $\hat{S}_w(s)$ is inferred from a set of observed data pairs, experimental measurements can exhibit substantial variability in both the range and scale of suction pressure $s$ and water saturation ratio $S_w$, depending on the type of porous material. 
Directly utilizing such raw data may compromise the stability and efficiency of machine learning models. 
To address this, we preprocess each data pair by mapping $(s, S_w)$ to normalized coordinates $(s^*, S^*_w)$, using reference points $(s_\text{min}, S_{w,\text{max}})$ and $(s_\text{res}, S_{w,\text{res}})$. 
Specifically, the suction pressure $s$ is first subjected to a logarithmic transformation and then linearly mapped to $s^* \in [0, 1]$, while the water saturation ratio $S_w$ is normalized based on its minimum and maximum values to yield $S^*_w \in [0, 1]$, as illustrated in Figure~\ref{fig:training_space}. 
Here, $S_{w,\text{max}}$ and $S_{w,\text{res}}$ represent the maximum and residual saturation levels, respectively, while $s_\text{min}$ and $s_\text{res}$ correspond to the minimum suction required to initiate air phase invasion and the suction at which residual is reached. 
Utilizing this mapping, the data loss term in the mapped space can be expressed as,
\begin{equation}
\label{eq:eq_data_loss_mapped}
\mathcal{L}_\text{data} = \frac{1}{n_\text{data}} \sum_{i=1}^{n_\text{data}} \left[ {\hat{S}_w}^{*(i)}(s^{*(i)}) - S^{*(i)}_w \right]^2,
\end{equation}
while the shape-guiding loss component can be determined as,
\begin{equation}
\label{eq:eq_mode_loss_mapped}
\mathcal{L}_\text{mode} = \left( \hat{N}_\text{mode} - N_\text{mode} \right)^2,
\end{equation}
noting that $N_\text{mode}$ is independent to the mapping. 
In this work, all training is conducted in the mapped space, with $S_{w,\text{max}}$ set to 1 to ensure that the drying process begins from a fully saturated condition.
It should be noted that these mapping parameters are auxiliary quantities determined by the observed data range to guide the search process, rather than intrinsic material parameters.

Small perturbations in mapping parameters or data can influence the form of the discovered equations, especially given the stochastic nature of the genetic programming-based symbolic regression. Nonetheless, it should be noted that the primary goal of this work is not to identify a unique ground-truth symbolic expression, but rather to discover a closed-form expression that best represents the given dataset with rigorously satisfying imposed physics constraints. This approach ensures physically consistent and interpretable models, despite inevitable variations in symbolic form.

\subsection{Physics constraints for water retention curves}
\label{sec:swrc_physics}
Ensuring that the learned functions are physically consistent and applicable to real porous systems requires satisfying a set of fundamental constraints grounded in thermodynamic principles. 
These constraints define the essential conditions that any valid water retention model should satisfy when characterizing the main drying path. 
Assuming that the learned function is continuous, the following outlines these critical conditions:

\paragraph*{Monotonicity constraint.}
A key requirement for the drying water retention curve is that as $s$ increases, $S_w$ must decrease. 
In other words, the degree of saturation should be a monotonically decreasing function of matric suction: 
\begin{equation}
\label{eq:eq_mono}
\frac{\mathrm{d} S_w}{\mathrm{d} s} \le 0. 
\end{equation}
This ensures that water is not spontaneously added to the pore space as matric suction increases, which would contradict thermodynamic principles.

\paragraph*{Limiting constraints.}
The drying path of a water retention curve must exhibit specific limiting behaviors at both ends of the matric suction spectrum. 
At $s \le s_\text{min}$, the degree of saturation should remain constant at its maximum value, $S_w = S_{w,\text{max}}$. 
Furthermore, the rate of change of saturation should approach zero at this wet end, i.e., 
\begin{equation}
\label{eq:eq_init}
S_w(s_\text{min}) = S_{w,\text{max}}
\: \: ; \: \:
\left. \frac{\mathrm{d} S_w}{\mathrm{d} s} \right|_{s = s_\text{min}} = 0.
\end{equation}
Conversely, under extremely dry conditions ($s \ge s_\text{res}$), the water saturation should approach a residual level as most pores have drained and only a small amount of water remains. 
This means that, at $s = s_\text{res}$, the saturation approaches its residual value and the rate of change of $S_w$ should approach 0:
\begin{equation}
\label{eq:eq_res}
S_w(s_\text{res}) = S_{w,\text{res}}
\: \: ; \: \:
\left. \frac{\mathrm{d} S_w}{\mathrm{d} s} \right|_{s = s_\text{res}} = 0.
\end{equation}

\paragraph*{Boundedness constraint.}
Regardless of the positioning of the reference points, the suction-saturation relationship must be bounded within physical limits. 
Specifically, the degree of saturation $S_w$ must satisfy:
\begin{equation}
\label{eq:eq_bound}
0 \le S_w \le 1.
\end{equation}
This constraint ensures that the saturation does not exceed unity nor fall below zero, aligned with the concept that the maximum water content corresponds to a fully saturated condition, while preventing non-physical scenarios, such as negative water content. 

Since we train our model in the mapped space, the physics constraints [Eqs.~\eqref{eq:eq_mono}-\eqref{eq:eq_bound}] must be reformulated accordingly. 
By letting the prime symbol denote the derivative with respect to the normalized suction $s^*$, the monotonicity constraint in the mapped space becomes:
\begin{equation}
\label{eq:eq_mono_mapped}
{S^*_w}'(s^*) 
 \le 0,  
\end{equation}
since both $\mathrm{d} S^*_w/\mathrm{d} S_w$ and $\mathrm{d} s/\mathrm{d} s^*$ are non-negative. 
Similarly, the limiting constraints at the wet end in the mapped space can be expressed as:
\begin{equation}
\label{eq:eq_init_mapped}
S^*_w(0) = 1
\: \: ; \: \:
{S^*_w}'(0) = 0,
\end{equation}
while the corresponding constraints at the dry end are given by,
\begin{equation}
\label{eq:eq_res_mapped}
S^*_w(1) = 0
\: \: ; \: \:
{S^*_w}'(1) = 0. 
\end{equation}
It is worth noting that the boundedness constraint is automatically satisfied in the mapped space if the saturation levels at the two reference points satisfy Eq.~\eqref{eq:eq_bound}, owing to the other constraints. 
This means that the boundedness constraint does not need to be explicitly enforced during model training.

\subsection{A physics-constrained symbolic regressor for closed-form water retention model discovery}
\label{sec:pcsr}
\begin{figure}[htbp]
\centering
\includegraphics[height=0.275\textwidth]{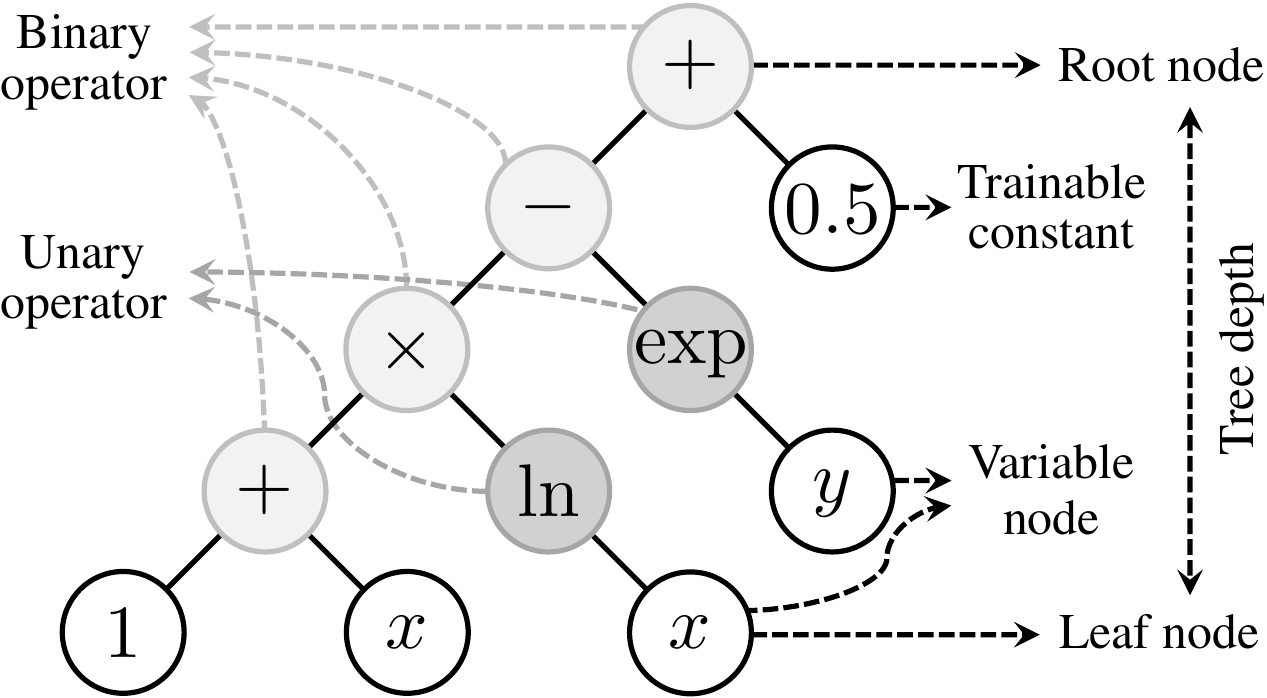}
\caption{Exemplary symbolic equation represented as a binary tree. This tree represents an expression $(1 + x)\ln(x) - \exp(y) + 0.5$, with a depth of 4 and a total size of 12 nodes.}
\label{fig:bin_tree}
\end{figure}

Symbolic regression aims to discover an explicit mathematical expression that best fits a given dataset, without prescribing its form in advance. 
The search process typically involves exploring a large space of candidate expressions, each represented as a binary tree, constructed from a predefined set of variables, constants, and mathematical operators (see Figure~\ref{fig:bin_tree}). 
Here, the total number of nodes in each discovered expression reflects its complexity, which, together with accuracy, serves as a key metric for evaluating model performance. 
In this work, a genetic programming framework \citep{koza1994genetic, wagner2005heuristiclab, schmidt2009distilling} is employed to navigate this space, where evolutionary operations are applied to iteratively refine the population of candidates toward increased predictive accuracy and physical consistency. 

\begin{figure}[htbp]
\centering

\begin{subfigure}[b]{0.37\textwidth}
  \centering
  \includegraphics[height=5.9cm,keepaspectratio,max width=\linewidth]{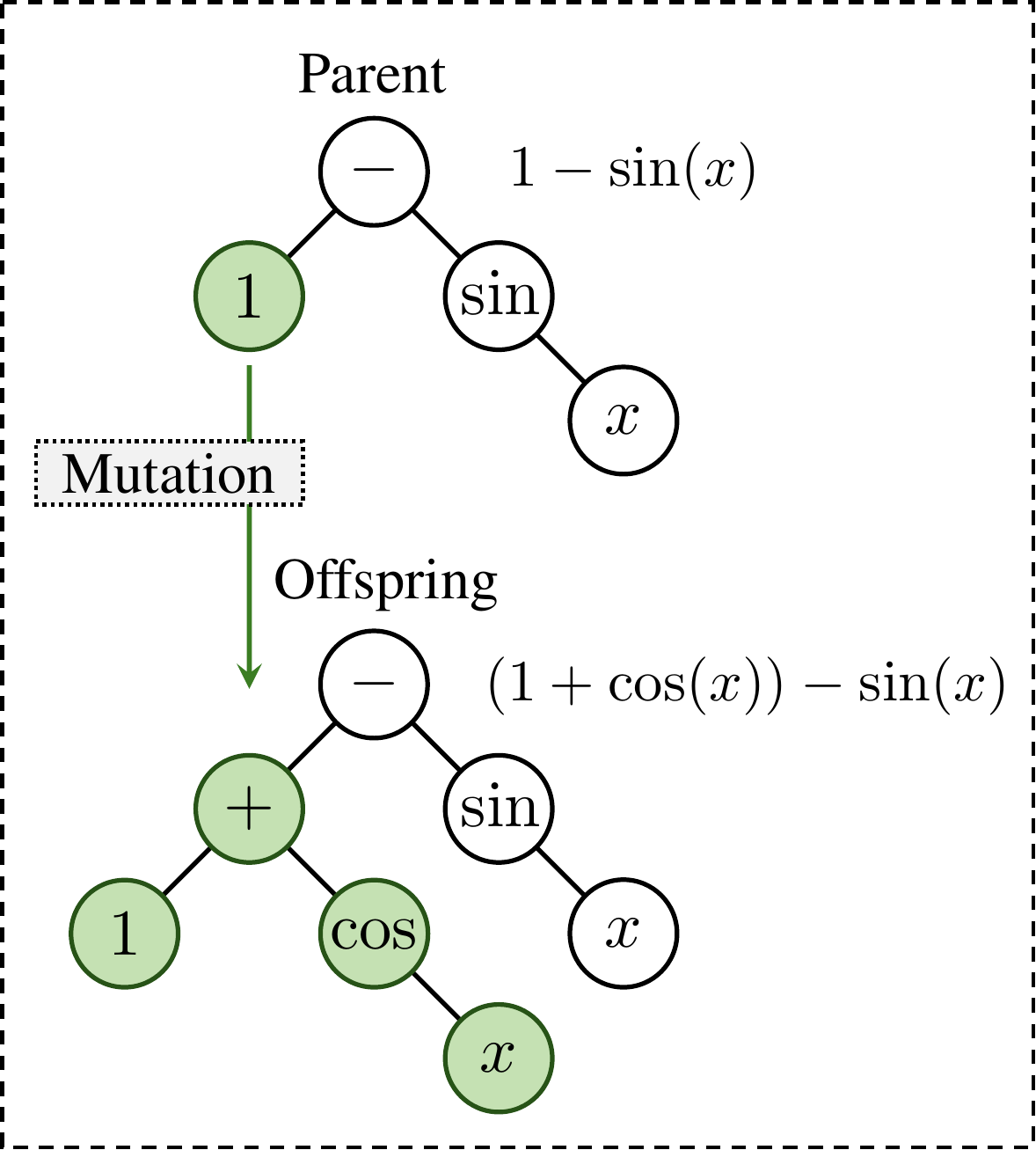}
  \caption{}
  \label{fig:mutation}
\end{subfigure}
\hfill
\begin{subfigure}[b]{0.61\textwidth}
  \centering
  \includegraphics[height=6cm,keepaspectratio,max width=\linewidth]{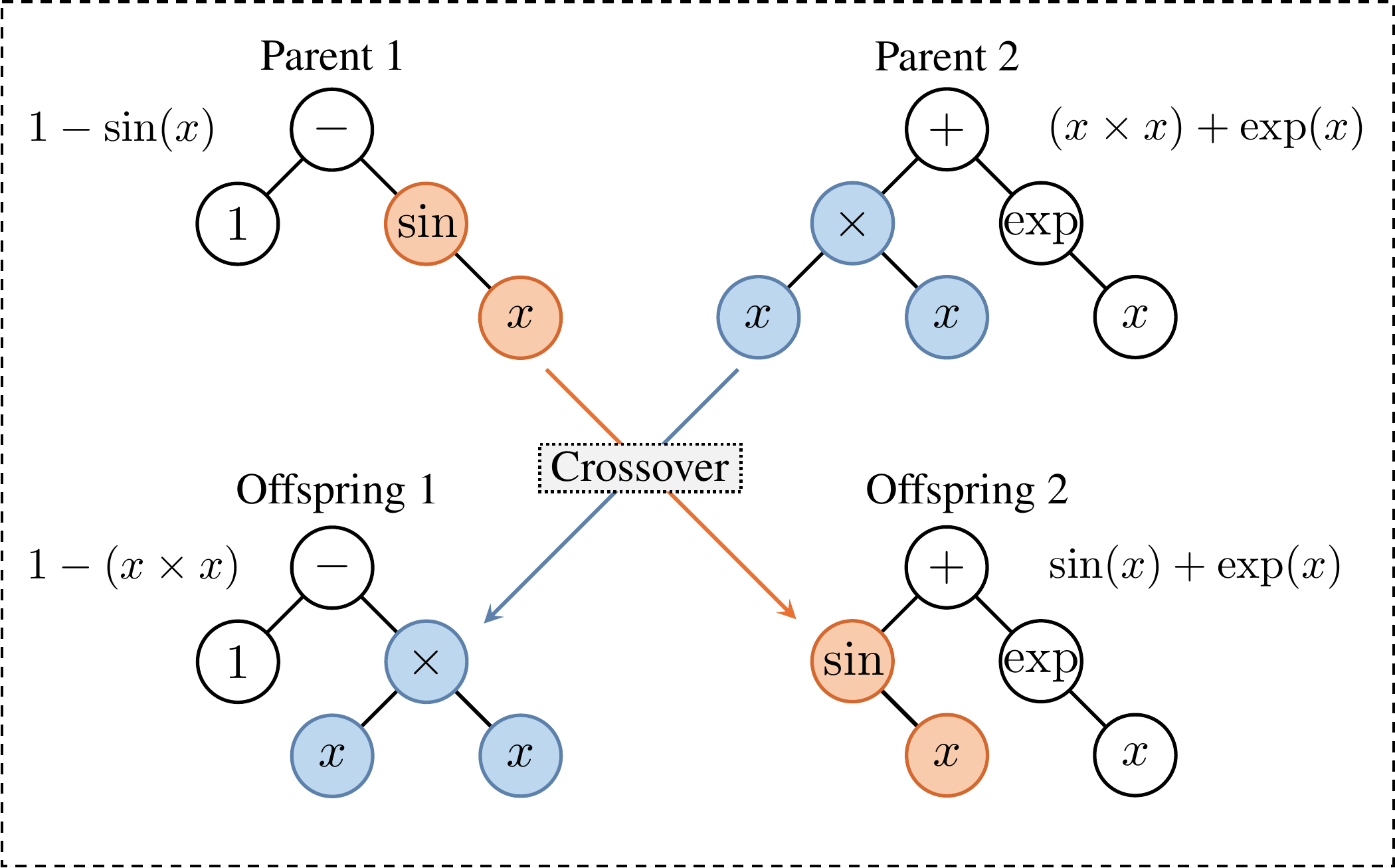}
  \caption{}
  \label{fig:crossover}
\end{subfigure}

\caption{Illustration of (a) mutation and (b) crossover operations in an evolutionary symbolic regression algorithm.}
\label{fig:genetic_algo}
\end{figure}

The training starts with randomly generating an initial population of candidate expressions. 
This population then undergoes an iterative evolutionary process consisting of three main operations: selection, mutation, and crossover. 
Selection involves choosing individuals with better performance from the current populations. 
Mutation, as shown in Figure~\ref{fig:mutation}, introduces small, random alterations to individual candidates, promoting diversity within the population. 
Crossover, illustrated in Figure~\ref{fig:crossover}, combines genetic information from two individuals to generate offspring that inherit characteristics from both parents. 

While vanilla symbolic regression (SR) aims solely to minimize the discrepancy between model predictions and data [e.g., using the mean squared error (MSE) loss], this study extends this approach by incorporating physics-based constraints outlined in Section~\ref{sec:swrc_physics} and a priori mode information into the learning process. 
In our PCSR, the standard MSE loss is replaced with a custom loss function that explicitly incorporates both physics and mode constraints. 
Specifically, each constraint is incorporated as an individual loss term that quantifies the discrepancy between the enforced constraint and the model’s behavior, and these terms are integrated to the total loss function as a penalty to guide the search process during training. 

\begin{algorithm}
\caption{Custom loss function for physics-constrained symbolic water retention model discovery.}
\label{alg:custom_loss}
\begin{algorithmic}
\Require Symbolic expression tree in the mapped space $\hat{S}^*_w(s)$, normalized dataset $\lbrace (s^{*(i)}, S_w^{*(i)}) \rbrace_{i=1}^{n_\text{data}}$, the prescribed number of modes $N_\text{mode}$, and collocation points $\vec{s}_\text{col}^* = \lbrace s_{\text{col}}^{*(j)} \rbrace_{j=1}^{n_\text{col}}$. 
\vspace{0.01\textwidth}
\Ensure Total loss $\mathcal{L}$.
\vspace{0.01\textwidth}
\State \textbf{1.} Evaluate $\mathcal{L}_\text{data}$. 
\Indent
\State $\mathcal{L}_\text{data} \leftarrow \frac{1}{n_\text{data}} \sum_{i=1}^{n_\text{data}} \left[ {\hat{S}_w}^{*(i)}(s^{*(i)}) - S^{*(i)}_w \right]^2$ \Comment{Eq.~\eqref{eq:eq_data_loss_mapped}}
\EndIndent
\State \textbf{2.} Evaluate $\mathcal{L}_\text{phys}$.
\Indent
\If{$\text{any}({\hat{S}^*}_w{'}(\vec{s}_\text{col}^*) > 0)$} 
\State $\mathcal{L}_\text{phys} \leftarrow \mathcal{L}_\text{phys} + \frac{1}{\text{nnz}\left({\hat{S}^*}_w{'}(\vec{s}_\text{col}^*) > 0\right)} \sum_{j=1}^{n_\text{col}} \left[ \text{ReLU} \left( {\hat{S}^*}_w{'}(s_\text{col}^{*(j)}) \right) \right]^2$ \Comment{Eq.~\eqref{eq:eq_mono_mapped}}
\Else
\State $\mathcal{L}_\text{phys} \leftarrow 0$
\EndIf
\If{$|\hat{S}^*_w(0) | < 1$} 
\State $\mathcal{L}_\text{phys} \leftarrow \mathcal{L}_\text{phys} + \left[ \hat{S}^*_w(0) - 1 \right]^2$ \Comment{Eq.~\eqref{eq:eq_init_mapped}$_1$}
\EndIf
\If{$|{\hat{S}^*}_w{'}(0)| < 0$} 
\State $\mathcal{L}_\text{phys} \leftarrow \mathcal{L}_\text{phys} + \left[{\hat{S}^*}_w{'}(0) \right]^2$ \Comment{Eq.~\eqref{eq:eq_init_mapped}$_2$}
\EndIf
\If{$|\hat{S}^*_w(1) | < 0$} 
\State $\mathcal{L}_\text{phys} \leftarrow \mathcal{L}_\text{phys} + \left[\hat{S}^*_w(1) \right]^2$ \Comment{Eq.~\eqref{eq:eq_res_mapped}$_1$}
\EndIf
\If{$|{\hat{S}^*}_w{'}(1) | < 0$} 
\State $\mathcal{L}_\text{phys} \leftarrow \mathcal{L}_\text{phys} + \left[\hat{S}^*_w{'}(1) \right]^2$ \Comment{Eq.~\eqref{eq:eq_res_mapped}$_2$}
\EndIf
\EndIndent
\State \textbf{3.} Evaluate $\mathcal{L}_\text{mode}$.
\Indent
\vspace{0.01\textwidth}
\State $\hat{N}_\text{mode} \leftarrow \text{nummodes}({\hat{S}^*}_w(\vec{s}_\text{col}^*))$
\vspace{0.01\textwidth}
\IfNot{$\hat{N}_\text{mode} == N_\text{mode}$}
\vspace{0.01\textwidth}
\State $\mathcal{L}_\text{mode} \leftarrow \left( \hat{N}_\text{mode} - N_\text{mode} \right)^2$ \Comment{Eq.~\eqref{eq:eq_mode_loss_mapped}}
\Else
\State $\mathcal{L}_\text{mode} \leftarrow 0$
\EndIf
\EndIndent
\State \textbf{4.} Return total loss $\mathcal{L}$. 
\Indent
\Return $\mathcal{L} \leftarrow \mathcal{L}_\text{data} + \mathcal{L}_\text{phys} + \mathcal{L}_\text{mode}$ \Comment{Eq.~\eqref{eq:loss_func}}
\EndIndent
\end{algorithmic}
\end{algorithm}

Similar to training physics-informed neural networks \citep{raissi2019physics, cai2021physics, haghighat2021physics}, we introduce a total of $n_\text{col}$ collocation points $\vec{s}_\text{col}^* = \lbrace s_{\text{col}}^{*(j)} \rbrace_{j=1}^{n_\text{col}}$, which are distributed uniformly across the mapped domain and serve as specific locations where constraints are explicitly enforced. 
As detailed in Algorithm~\ref{alg:custom_loss}, the custom loss function is constructed by first evaluating the data loss ($\mathcal{L}_\text{data}$) on the normalized dataset to ensure fitness to experimental observations. 
Subsequently, the physics loss ($\mathcal{L}_\text{phys}$) is computed by checking the predictions at the collocation points for violations of physics constraints. 
Monotonicity constraint is enforced by penalizing deviations from Eq.~\eqref{eq:eq_mono_mapped}: at each collocation point where the monotonicity condition is not met, a penalty term $[ \text{ReLU} ( {\hat{S}^*}_w{'}(s_\text{col}^{*(j)})) ]^2$ is added to the physics loss. 
Here, $\text{ReLU}( \bullet )$ denotes the rectified linear unit, which ensures that only instances where the derivative exceeds zero are penalized, thereby quantifying the extent of non-monotonicity. 
Limiting constraints are similarly incorporated by imposing additional penalties for deviations from the expected values and derivatives at the endpoints of the domain ($s^*=0$ and $s^*=1$), thus ensuring physical consistency at both maximum and residual saturation levels. 
The mode loss ($\mathcal{L}_\text{mode}$) is then evaluated using Eq.~\eqref{eq:eq_mode_loss_mapped}. 
Here, $\text{nummodes}( {\hat{S}^*}_w(\vec{s}_\text{col}^*) )$ determines the number of modes of the learned function ($\hat{N}_\text{mode}$), by counting the number of sign changes from negative to positive in the second derivative ${\hat{S}^*_w}{''}$, which corresponds to the number of concave-to-convex inflection points. 
Finally, the custom loss function returns the total loss $\mathcal{L}$ as defined in Eq.~\eqref{eq:loss_func}.

To further investigate how the imposed physics constraints suppress nonphysical solutions and effectively mitigate overfitting, particularly in the presence of noisy data, we provide a detailed comparative analysis between vanilla SR and the constrained variants in \nameref{sec:apd}.

The entire PCSR implementation is built on top of the Julia package \verb|SymbolicRegression.jl| \citep{cranmer2023interpretable}, and is open-sourced to support third-party application and extension (see \nameref{sec:data_avail}).

\section{Results}
\label{sec:results}
This section presents a comprehensive evaluation of the proposed PCSR framework for discovering closed-form equations of water retention models under various scenarios. 
To establish a thorough benchmark, we compare the performance of PCSR against both the vanilla SR and established semi-empirical models \citep{van1980closed, durner1994hydraulic}. 
In Section~\ref{sec:unimodal}, we apply PCSR to two distinct experimental datasets representing soils with unimodal pore size distributions and examine its ability to discover physically consistent equations. 
Section~\ref{sec:multimodal} extends the evaluation to multimodal water retention curves, including both experimentally measured bimodal  ($N_\text{mode}=2$) datasets and synthetically generated datasets with higher mode numbers (e.g., $N_\text{mode}=3$ and $4$), to investigate the scalability and robustness of PCSR in capturing complex, multi-scale retention behaviors as well as its effectiveness in enforcing the prescribed number of modes.

All symbolic regressions in this work were trained using a consistent set of hyperparameter settings: the number of training iterations was fixed at 100, the population size of candidate expressions was set to 50, and the maximum allowable complexity for each candidate expression was limited to 150 nodes. 
The binary operators available for the model were restricted to $\lbrace +, \times \rbrace$, while the unary operators included $\lbrace \sin( \bullet ), \cos( \bullet ), \exp( \bullet ), \ln( \bullet ) \rbrace$. 
For PCSRs, the number of collocation points was fixed at $n_\text{col} = 100$. 

It is important to note that a conventional train-test splitting strategy is not adopted in this study. The objective of the proposed framework is not to construct a model that generalizes across different datasets, but rather to discover a physically sound, best-fit closed-form equation specific to the given experimental dataset (i.e., meta-modeling). Hence, all available measurements are fully leveraged as training data, while overfitting is effectively controlled by the imposed physical constraints, which limit the admissible functional space and prevent nonphysical solutions that might otherwise fit local noise or fluctuations.  Furthermore, because our primary focus is on ensuring that the discovered equations satisfy the physical constraints rather than preventing overparameterization (e.g., minimizing expression complexity), the resulting equations may be highly complex, yet remain physically consistent.

\subsection{Unimodal water retention curves}
\label{sec:unimodal}
We considered two different water retention datasets: one corresponding to the poorly graded sand, which exhibits a steep slope in its water retention curve \citep{song2014suction}, and the other to the clay loam soil with a water retention curve \citep{tuller2001hydraulic}. 
For each case, the experimental data were mapped to the normalized space as described in Section~\ref{sec:learn_prob} for training, using the following mapping parameters: for the poorly graded sand, $s_\text{min}=10^{-2}$, $S_{w,\text{max}}=1.0$, and $s_\text{res}=10^{5}$; for the clay loam soil, $s_\text{min}=10^{-1}$, $S_{w,\text{max}}=1.0$, $s_\text{res}=10^{6}$, and $S_{w,\text{res}}=0.035$. 
Model predictions were then obtained by applying the inverse mapping to the outputs of the trained models.

\begin{figure}[htbp]
\centering

\begin{subfigure}[b]{0.49\textwidth}
  \centering
  \includegraphics[width=\linewidth]{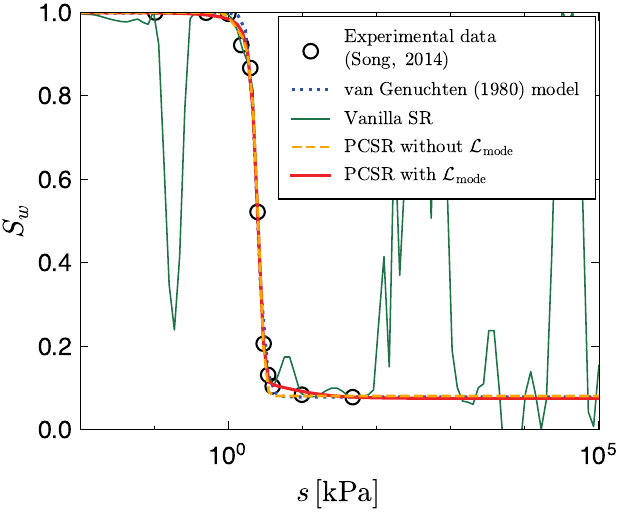}
  \caption{}
  \label{fig:uni_a}
\end{subfigure}
\hfill
\begin{subfigure}[b]{0.49\textwidth}
  \centering
  \includegraphics[width=\linewidth]{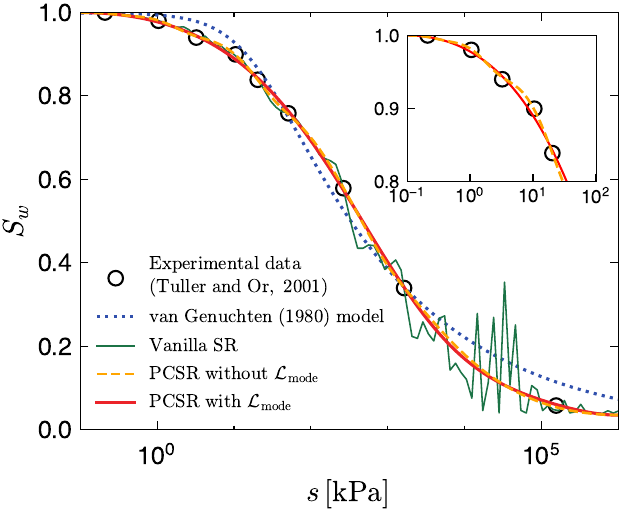}
  \caption{}
  \label{fig:uni_b}
\end{subfigure}

\caption{The symbolic model discovered from experimental unimodal water retention data using the proposed approach, compared against the \citep{van1980closed} model, vanilla SR, and PCSR without mode constraint.}
\label{fig:unimodal_wrc}
\end{figure}

Figure~\ref{fig:unimodal_wrc} presents the water retention models discovered using PCSR with mode constraint (red solid curves) and without mode constraint (yellow dashed curves), as well as those obtained with vanilla SR (green solid curves) and the \citep{van1980closed} model (blue dotted curves). 
Here, the experimental training data are depicted as black hollow circles. 
While the semi-empirical model closely fits the experimental data for poorly graded sand (Figure~\ref{fig:uni_a}), it may fail to accurately capture the retention behavior of the clay loam soil near the air entry value or residual saturation (Figure~\ref{fig:uni_b}). 
Although the models identified by vanilla SR achieve very low MSE losses and closely match the experimental data, they tend to overfit and exhibit limited generalizability beyond the training dataset. 
When physics loss $\mathcal{L}_\text{phys}$ is incorporated into the vanilla SR (i.e., PCSR without $\mathcal{L}_\text{mode}$), excessive fluctuations are effectively suppressed, ensuring that the resulting curves adhere to the physics constraints detailed in Section~\ref{sec:swrc_physics}. 
However, as depicted in the subfigure within Figure~\ref{fig:uni_b}, the curve may still exhibit slight overfitting, which does not represent a unimodal pore size distribution. 
In contrast, the proposed framework that includes both physics and mode constraints (i.e., PCSR with $\mathcal{L}_\text{mode}$), takes the target number of modes as an additional input (e.g., $N_\text{mode} = 1$ in this case), and effectively prevents overfitting to the sparse experimental data, resulting in water retention models that only exhibit one concave-to-convex inflection point. 

Table~\ref{tab:uni_fit_eqs} summarizes the best-fit mathematical expressions discovered for each unimodal water retention dataset using PCSR, both with and without mode constraints, corresponding to those illustrated in Figures \ref{fig:uni_a}-\ref{fig:uni_b}. 
For each case, we also provide the number of modes exhibited by the trained model ($\hat{N}_\text{mode}$), the complexity of the symbolic expression, and the corresponding total loss $\mathcal{L}$. 
Models trained without the mode loss term all fail to achieve $\hat{N}_\text{mode} = 1$, whereas models incorporating $\mathcal{L}_\text{mode}$ tend to consistently achieve $\hat{N}_\text{mode}=1$, confirming that the identified water retention curve is unimodal. 
Further, the best-fit mathematical expressions exhibit sufficiently low total loss values, and complexities below the maximum bound of 150, indicating that the proposed approach can efficiently identify concise, physically sound water retention curves with a consistent number of modes for unimodal datasets. 

\begin{sidewaystable}
\centering
\caption{Discovered closed-form equations corresponding to those illustrated in Figures \ref{fig:uni_a}--\ref{fig:uni_b}.}
\label{tab:uni_fit_eqs}

\renewcommand{\arraystretch}{1.25}

\begin{tabular}{lllccc}
\toprule
& \textbf{Model}
& \textbf{Discovered expression}
& \textbf{$\hat{N}_\text{mode}$}
& \textbf{Complexity}
& \textbf{Loss} \\
\midrule

(a)
& PCSR without $\mathcal{L}_\text{mode}$
& {\small
$\begin{aligned}
S_w &= \sin( \sin( ( \cos( \cos( \exp(\exp(( \sin( s ) -0.32 ) \times 64.88) \times -0.25) ) ) + 0.85 ) \\
&\times 4.51 ) \times ( \cos( \cos( \exp(\exp(( \sin( s ) -0.27 ) \times 64.88) \times -0.25) ) ) + 0.76 ) ) \times 1.00
\end{aligned}$}
& 2
& 35
& $3.751 \times 10^{-4}$ \\

\addlinespace[1.2ex]

{}
& PCSR with $\mathcal{L}_\text{mode}$
& {\small
$\begin{aligned}
S_w &= \sin( \sin( \exp(\exp(s \times 97.45 \times s) \times -1.46\text{E}{-05} + 0.25) ) \times 1.44 ) \times 0.94 \\
&+ \exp(\cos( ( \sin( s + s ) + s ) \times ( s + \sin( s + s ) ) \times \sin( s + s ) ) \times 2.69) \times 0.005
\end{aligned}$}
& 1
& 42
& $2.198 \times 10^{-4}$ \\

\midrule

(b)
& PCSR without $\mathcal{L}_\text{mode}$
& {\small
$\begin{aligned}
S_w &= \cos( \cos( \sin( \sin( \cos( s \times ( s -2.12 ) \times -12.45 ) \times s ) \times -1.70 ) ) \\
&\times 0.46 ) \times ( \sin( \cos( \cos( s \times -3.14 ) + 4.75 ) ) \times 0.75 + 0.55 )
\end{aligned}$}
& 4
& 31
& $4.790 \times 10^{-6}$ \\

\addlinespace[1.2ex]

{}
& PCSR with $\mathcal{L}_\text{mode}$
& {\small
$\begin{aligned}
S_w &= \cos( \sin( ( ( s + 0.64 ) \times \sin( s \times 1.88 ) \times 1.42 + s ) \\
&\times ( s -0.23 ) ) ) \times \cos( \sin( s \times 1.88 ) \times ( s + 0.65 ) )
\end{aligned}$}
& 1
& 28
& $3.023 \times 10^{-5}$ \\

\bottomrule
\end{tabular}
\end{sidewaystable}

\begin{figure}[htbp]
\centering

\begin{subfigure}[b]{0.49\textwidth}
  \centering
  \includegraphics[width=\linewidth]{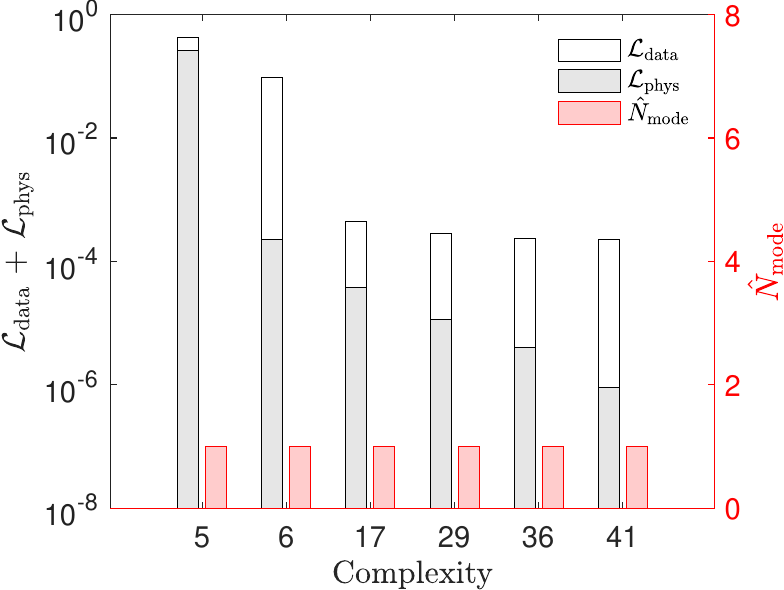}
  \caption{}
  \label{fig:uni_comp_loss_a}
\end{subfigure}
\hfill
\begin{subfigure}[b]{0.49\textwidth}
  \centering
  \includegraphics[width=\linewidth]{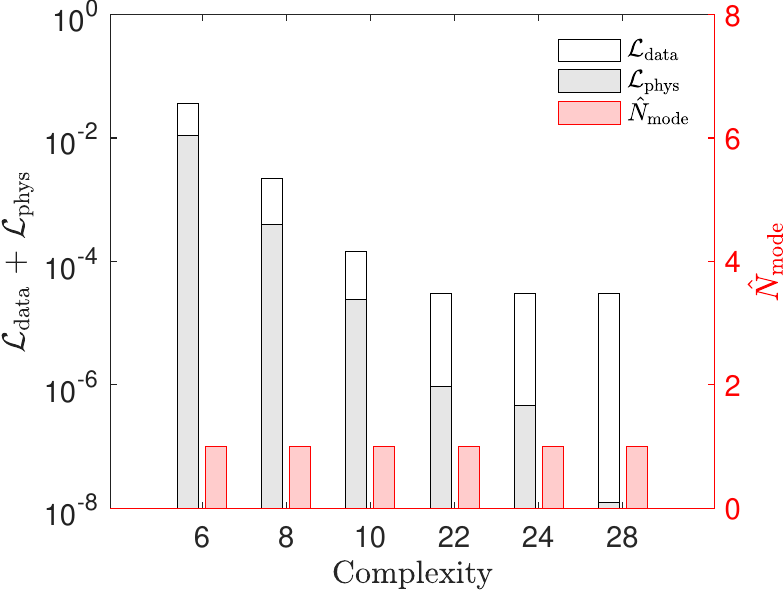}
  \caption{}
  \label{fig:uni_comp_loss_b}
\end{subfigure}

\caption{Data and physics loss components and the number of modes exhibited by PCSR with $\mathcal{L}_\text{mode}$ at varying complexity levels that optimally fit the data presented in Figure~\ref{fig:unimodal_wrc}.}
\label{fig:uni_complex_loss}
\end{figure}

Figure~\ref{fig:uni_complex_loss} showcases six exemplary mathematical expressions discovered by the PCSR with $\mathcal{L}_\text{mode}$, each exhibiting varying levels of complexity, illustrating its effect on the loss and the number of modes exhibited. 
Here, the data ($\mathcal{L}_\text{data}$) and physics ($\mathcal{L}_\text{phys}$) losses are represented as white and gray segments in the stacked bar plots on the left primary y-axis, respectively, while the exhibited number of modes $\hat{N}_\text{mode}$ as a red bar on the right secondary y-axis. 
The results suggest that as the level of complexity increases, both $\mathcal{L}_\text{data}$ and $\mathcal{L}_\text{phys}$ tend to decrease, indicating that the discovered mathematical expression better describes the water retention behavior as the number of nodes in the binary tree increases. 
This highlights that there is a trade-off between accuracy and simplicity.   
Further, when $\mathcal{L}_\text{mode}$ is incorporated into the PCSR, mathematical expressions consistently maintain $\hat{N}_\text{mode}=1$, demonstrating that the mode loss term effectively enforces the intended unimodal shape in the discovered water retention curves.  

\begin{figure}[htbp]
\centering

\begin{subfigure}[b]{\textwidth}
  \centering
  \includegraphics[width=\textwidth]{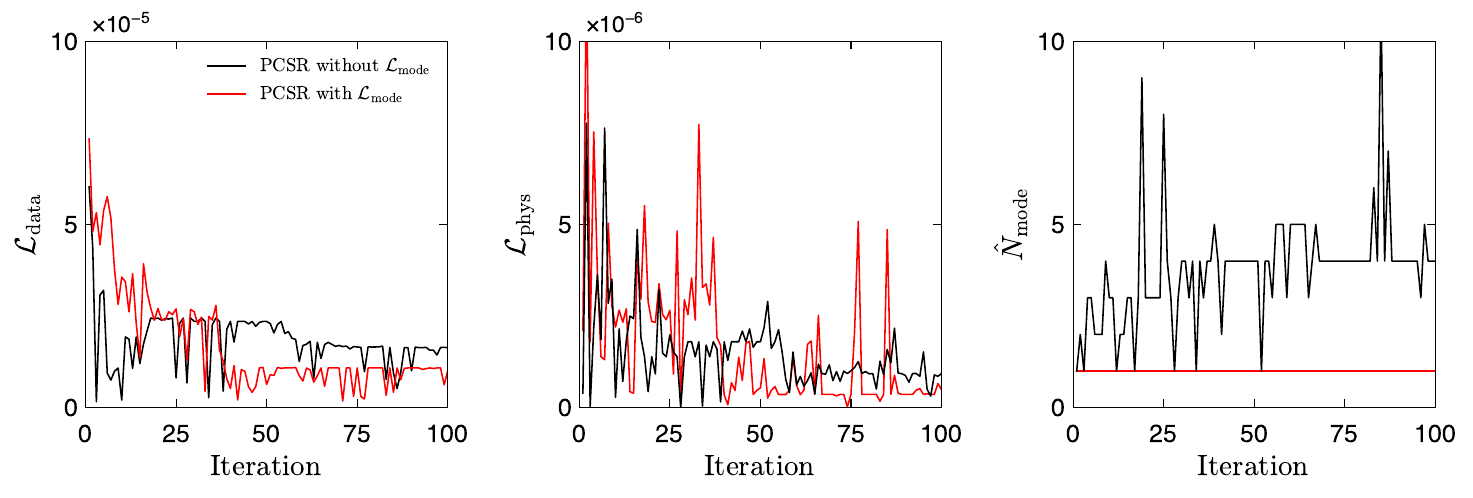}
  \caption{}
  \label{fig:uni_iter_loss_a}
\end{subfigure}

\vspace{0.6em}

\begin{subfigure}[b]{\textwidth}
  \centering
  \includegraphics[width=\textwidth]{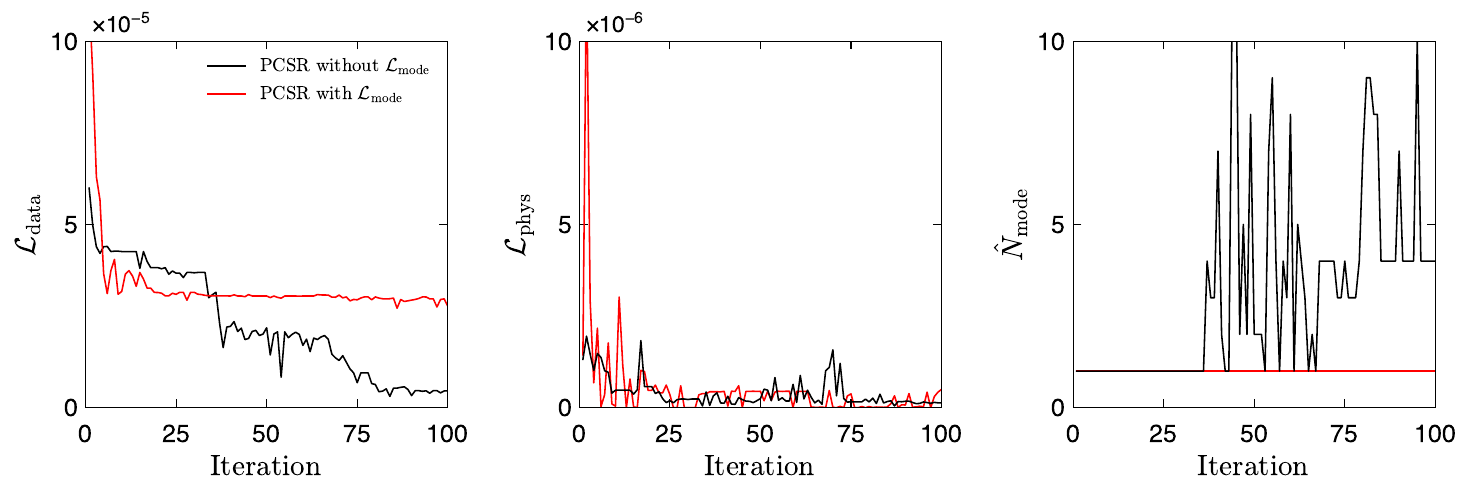}
  \caption{}
  \label{fig:uni_iter_loss_b}
\end{subfigure}

\caption{Learning curves of loss components and predicted number of modes for the PCSR without and with $\mathcal{L}_\text{mode}$ for the data presented in Figure~\ref{fig:unimodal_wrc}.}
\label{fig:uni_iter_loss}

\end{figure}

We also compare the learning curves of PCSR models trained with and without $\mathcal{L}_\text{mode}$, as illustrated in Figure~\ref{fig:uni_iter_loss}. 
Figures~\ref{fig:uni_iter_loss_a} and \ref{fig:uni_iter_loss_b} each contain three subplots, displaying the evolution of $\mathcal{L}_\text{data}$, $\mathcal{L}_\text{phys}$, and, instead of $\mathcal{L}_\text{mode}$, the change in $\hat{N}_\text{mode}$ over training iterations. 
The black solid line represents the learning curve of PCSR without the mode loss term, while the red solid line indicates the results from PCSR with the shape-guiding mode loss term included. 
In both models, $\mathcal{L}_\text{data}$ and $\mathcal{L}_\text{phys}$ exhibit a downward trend as training progresses, though some noise is observed due to the evolutionary nature of symbolic regression. 
Nonetheless, the overall loss levels are low enough that the difference between the models is negligible. 
The difference between two models can be seen in the change of $\hat{N}_\text{mode}$ during training. 
PCSR trained without $\mathcal{L}_\text{mode}$ fails to achieve the target number of modes for the unimodal water retention curve, often increasing the expression complexity in an attempt to reduce data and physics losses. 
This results in overfitting and a substantial deviation of $\hat{N}_\text{mode}$ from the prescribed value, as shown in Figures~\ref{fig:uni_iter_loss_a} and \ref{fig:uni_iter_loss_b}. 
In contrast, the candidate representations obtained by PCSR with the mode loss term consistently yield $\hat{N}_\text{mode}=1$ throughout the training process, while simultaneously reducing the values of the other two loss components. 

These results demonstrate that incorporating both physics and mode loss terms into the symbolic regression training enables the discovery of closed-form mathematical expressions that can capture unimodal water retention behavior. 
Compared to other approaches, our PCSR avoids overfitting and spurious modes, consistently yielding physically sound models while maintaining the specified number of modes. 
Having demonstrated the effectiveness of the proposed approach for unimodal cases, we next consider its performance on more complex, multimodal water retention datasets.

\subsection{Multimodal water retention curves}
\label{sec:multimodal}
\begin{figure}[htbp]
\centering

\begin{subfigure}[b]{0.49\textwidth}
  \centering
  \includegraphics[width=\linewidth]{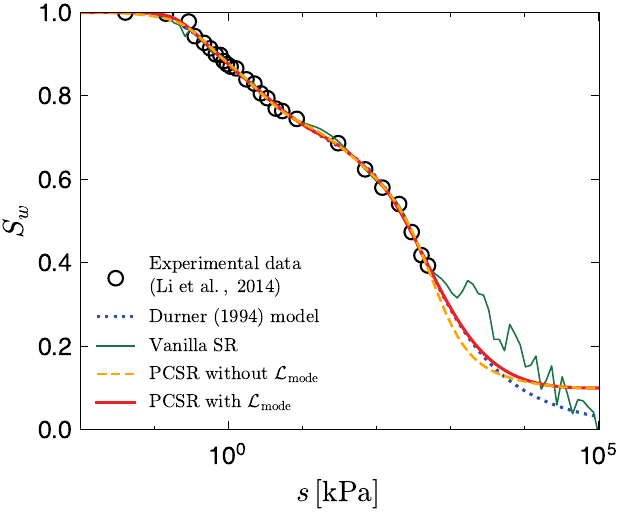}
  \caption{}
  \label{fig:bi_a}
\end{subfigure}
\hfill
\begin{subfigure}[b]{0.49\textwidth}
  \centering
  \includegraphics[width=\linewidth]{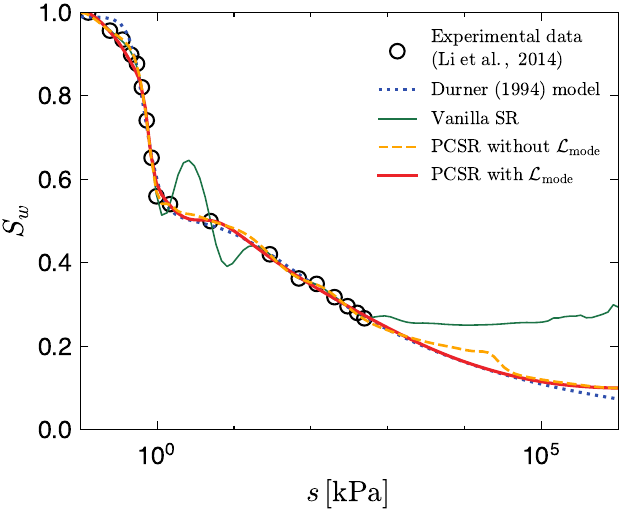}
  \caption{}
  \label{fig:bi_b}
\end{subfigure}

\vspace{0.5em}

\begin{subfigure}[b]{0.49\textwidth}
  \centering
  \includegraphics[width=\linewidth]{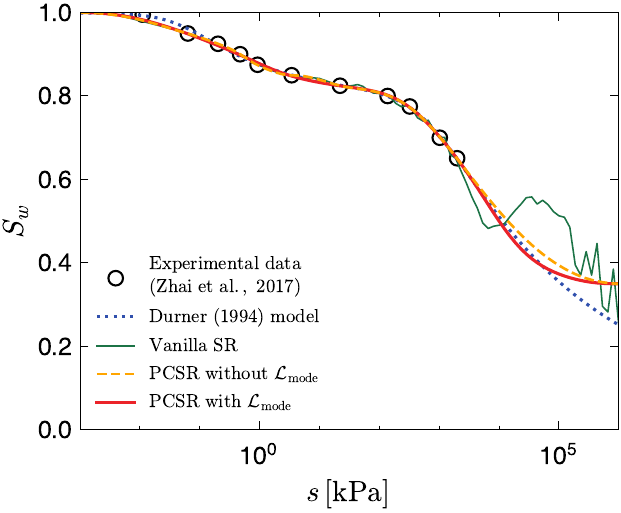}
  \caption{}
  \label{fig:bi_c}
\end{subfigure}
\hfill
\begin{subfigure}[b]{0.49\textwidth}
  \centering
  \includegraphics[width=\linewidth]{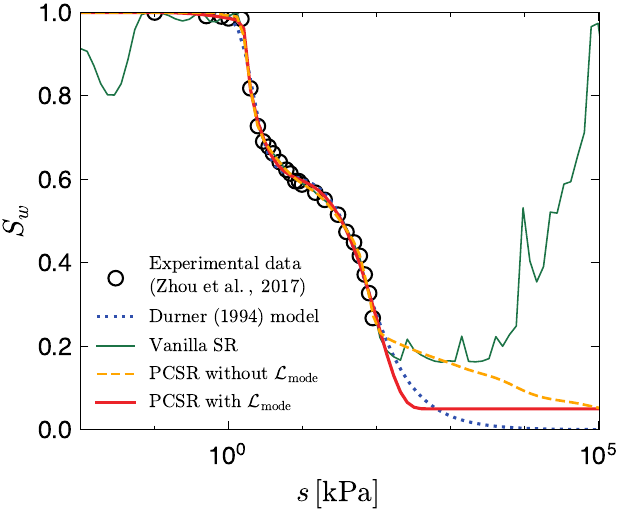}
  \caption{}
  \label{fig:bi_d}
\end{subfigure}

\caption{The symbolic model discovered from experimental bimodal water retention data using the proposed approach, compared against the \citep{durner1994hydraulic} model, vanilla SR, and PCSR without mode constraint.}
\label{fig:bimodal_wrc}

\end{figure}

In this section, we first consider the PCSR models trained with the bimodal water retention data ($N_\text{mode}=2$), utilizing four distinct training datasets that are collected from the literature. 
As shown in Figure~\ref{fig:bimodal_wrc}, these datasets were collected to test the scalability and applicability of our PCSR model, as they exhibit markedly different curve shapes and cover distinct suction pressure ranges. 
The datasets correspond to: (1) silty sand with gravel and (2) well-graded gravel with sand \citep{li2014predicting}, (3) paddy soil \citep{zhai2017effect}, and (4) a mixture of $90\%$ sand and $10\%$ kaolin \citep{zhou2017bimodal}. 
For symbolic regression model training, the mapping parameters for each dataset were set as follows: (1) $s_\text{min} = 10^{-2}$, $S_{w,\text{max}} = 1.0$, $s_\text{res} = 10^{5}$, and $S_{w,\text{res}} = 0.1$; (2) $s_\text{min} = 10^{-1}$, $S_{w,\text{max}} = 1.0$, $s_\text{res} = 10^{6}$, and $S_{w,\text{res}} = 0.1$; (3) $s_\text{min} = 10^{-3}$, $S_{w,\text{max}} = 1.0$, $s_\text{res} = 10^{6}$, and $S_{w,\text{res}} = 0.35$; and (4) $s_\text{min} = 10^{-2}$, $S_{w,\text{max}} = 1.0$, $s_\text{res} = 10^{5}$, and $S_{w,\text{res}} = 0.05$.

Figure~\ref{fig:bimodal_wrc} presents the best-fit water retention models discovered using PCSR, both with and without $\mathcal{L}_\text{mode}$, compared against the experimental results and the curve fitted using the \citep{durner1994hydraulic} model. 
It should be noted that the colors and styles of the symbols and curves used here are identical to those described previously in Figure~\ref{fig:unimodal_wrc}. 
The water retention curves discovered using vanilla SR (solid green curves) provide fairly accurate predictions from the given dataset, but substantial fluctuations are evident, similar to the unimodal results (Figure~\ref{fig:unimodal_wrc}). 
These issues may arise from focusing solely on minimizing the differences between the given data points and the model predictions, and can be effectively resolved by incorporating physics constraints, as demonstrated by the retention curves discovered by the PCSR without $\mathcal{L}_\text{mode}$ (solid yellow curves). 
Despite the improvements in curve fitting, as shown in Tables~\ref{tab:bi_fit_eqs_1} and \ref{tab:bi_fit_eqs_2}, the mathematical expressions discovered by PCSR without the mode loss term result in overestimated values of $\hat{N}_\text{mode}$--specifically 3, 9, 4, and 12 for the curves shown in Figures~\ref{fig:bi_a}–\ref{fig:bi_d}, respectively--and thus fail to match the targeted number of modes, $N_\text{mode}=2$. 
In contrast, the proposed framework (i.e., PCSR with $\mathcal{L}_\text{mode}$), successfully discovers the retention curves with the desired bimodal shapes, yielding $\hat{N}_\text{mode}=2$ and thereby demonstrating precise alignment with the target mode number. 

\begin{sidewaystable}
\caption{Discovered closed-form equations corresponding to those illustrated in Figures \ref{fig:bi_a}-\ref{fig:bi_b}.}
\label{tab:bi_fit_eqs_1}
\begin{tabular}{lllccc}
\hline
{}
& \textbf{Model}
& \textbf{Discovered expression}
& \textbf{$\hat{N}_\text{mode}$}
& \textbf{Complexity}
& \textbf{Loss}
\\
\hline
(a)
& PCSR without $\mathcal{L}_\text{mode}$
& {\small $
\begin{aligned}
	S_w &= \cos( \sin( \sin( \sin( s \times 1.01 + s + \sin( \sin( \exp(( s \times 2.72 + s ) \times 0.23) \\
	&+ \sin( s \times ( s + \sin( s ) + ( s \times 2.70 + s ) \times 0.27 + s + s + \sin( \sin( s ) \\
	&+ ( s + s + \sin( \sin( \sin( \sin( s + s + s \times ( s + 1.18 ) -0.51 ) ) + s + s + \sin( s ) \\
	&-2.24 + s ) + s + s \times s + s + s + \sin( s + s ) ) -0.53 ) \times \exp(( s \times 2.73 \\
	&+ s \times 0.27 ) \times -0.23 + 0.85) + s + s + s -1.59 ) -0.78 ) ) ) ) -1.41 ) \\
	&+ s -0.55 ) ) + \exp(\cos( ( s + 0.20 ) \times -0.57 ) + 1.38) \times 0.03 + 0.48 ) \times 1.00
\end{aligned}
$}
& {3}
& {137}
& {$4.555 \times 10^{-5}$}
\\
{}
& PCSR with $\mathcal{L}_\text{mode}$
& {\small $
\begin{aligned}
	S_w &= \sin( \exp(\sin( \exp(\cos( s \times \sin( s + s + s + 1.74 ) ) + 0.93) ) \\
	&+ \cos( \sin( \exp(\cos( \cos( \exp(\cos( \exp(\cos( \sin( s ) + \exp(\sin( s )) + s )) )) + s \\
	&+ \sin( \exp(\cos( \exp(\cos( \cos( \cos( \exp(\cos( \cos( s \times ( s -1.06 + s -0.19 ) ) \\
	&+ \exp(\cos( \sin( \exp(\cos( s \times ( s + 0.92 + s \times 1.94 ) ) + 0.92) ) )) + s )) ) \\
	&+ \cos( \cos( \sin( s ) + ( s + 0.68 ) \times s \times 1.97 ) ) ) + \exp(\cos( s + \sin( \sin( s ) ) \\
	&+ s + 0.50 )) + \exp(s \times s) )) )) + \exp(\cos( \sin( s ) ) + 0.86) ) ) ) \\
	&+ \sin( \exp(\cos( \sin( s \times ( s + 0.78 ) \times \cos( \sin( s + 2.95 ) ) ) ) + 0.88) )) ) \\
	&\times ( s + s \times 1.95 \backslash \mathrm{left}() ) ) -0.19) \times 0.35 )
\end{aligned}
$}
& {2}
& {141}
& {$4.407 \times 10^{-5}$}
\\
\hline
(b)
& PCSR without $\mathcal{L}_\text{mode}$
& {\small $
\begin{aligned}
	S_w &= \cos( \sin( \sin( s \times 1.55 ) + \sin( \sin( \sin( s ) ) \times 2.36 ) + ( \sin( \sin( \sin( s \times s \\
	&\times 1.68 + s ) + ( s + \sin( s ) + s ) \times 3.10 + s -1.02 ) ) + s ) \times ( \sin( ( \sin( s ) + \sin( s ) \\ 
	&+ s \times 0.97 + \sin( \sin( \sin( ( s + s + s ) \times 2.05 + s -1.29 ) + \sin( s -1.30 \\
	&+ ( \sin( s + s ) + 2.04 ) \times ( ( s + s + s ) \times 1.04 + s ) ) ) \times ( \sin( s + s + ( s + s \\
	&+ s + \sin( s ) ) \times 1.42 ) + s + \sin( \sin( s + s ) ) ) ) ) \times 1.36 + s + \sin( s + s \\
	&+ \sin( \sin( s ) ) ) ) + \sin( s ) + s ) ) ) \times \cos( ( \sin( \sin( s \times 1.02 + s ) ) + \sin( s ) ) \times 0.97 )
\end{aligned}
$}
& {9}
& {143}
& {$2.450 \times 10^{-5}$}
\\[2ex]
{}
& PCSR with $\mathcal{L}_\text{mode}$
& {\small $
\begin{aligned}
	S_w &= \cos( \cos( \sin( \sin( s \times 1.21 ) + \sin( \sin( s + s ) ) ) ) \times ( \sin( ( s + \sin( s \times 1.37 ) \\
	&+ \sin( s + s ) ) \times \cos( \sin( ( \sin( \sin( \sin( ( \sin( s ) + s + s -0.45 ) \times 1.07 ) \\
	&\times \exp(\sin( \sin( \sin( \sin( s + s + s -0.45 ) ) \times -1.70 ) \times \exp(\sin( \sin( \sin( ( s + s \\
	&+ 1.07 ) \times 3.96 ) + s \times 1.02 + \sin( s ) ) \times -1.68 ) + 1.04) ) + 0.99) + s ) + s ) \\
	&+ \sin( s ) ) \times -1.73 + \cos( \cos( s ) ) ) ) ) + \sin( \sin( s + s ) ) + \sin( s ) ) ) \times 1.20 -0.20
\end{aligned}
$}
& {2}
& {108}
& {$1.862 \times 10^{-4}$}
\\
\hline
\end{tabular}
\end{sidewaystable}

\begin{sidewaystable}
\caption{Discovered closed-form equations corresponding to those illustrated in Figures \ref{fig:bi_c}-\ref{fig:bi_d}.}
\label{tab:bi_fit_eqs_2}
\begin{tabular}{lllccc}
\hline
{}
& \textbf{Model}
& \textbf{Discovered expression}
& \textbf{$\hat{N}_\text{mode}$}
& \textbf{Complexity}
& \textbf{Loss}
\\
\hline
(c)
& PCSR without $\mathcal{L}_\text{mode}$
& {\small $
\begin{aligned}
	S_w &= \sin( \exp(\sin( \sin( \sin( \sin( s \times s ) ) \times ( s + \sin( \sin( s ) ) \times ( s + \sin( ( s + s ) \\
	&\times 1.12 ) ) ) ) )) -0.03 ) + \sin( s + \sin( \sin( \cos( s \times 1.69 ) \times s \\
	&\times ( s + \sin( \sin( s + s \times s + \sin( \sin( ( s + \sin( s ) + 1.48 ) \times 1.70 \\
	&\times ( \sin( s -0.55 ) + \exp(s) ) \times \cos( s ) \times ( \exp(\sin( s \times 1.00 + s + s )) -0.52 ) ) \\
	&+ \sin( s \times ( s + 4.81 ) ) \times -0.31 ) ) ) + 0.27 ) \times s \times ( \sin( \exp(\sin( s \times ( s + s \\
	&+ 1.41 ) )) ) -0.53 ) \times ( s + s + 1.43 ) + s ) ) ) \times \sin( s ) \times -0.91 ) \times 1.21
\end{aligned}
$}
& {4}
& {123}
& {$1.812 \times 10^{-5}$}
\\
{}
& PCSR with $\mathcal{L}_\text{mode}$
& {\small $
\begin{aligned}
	S_w &= \sin( ( \sin( s ) + ( s -2.02 ) \times \cos( ( \cos( ( s + s \times s \times \cos( s ) \times 0.63 ) \\
	&\times ( s \times -1.01 + \cos( \cos( s \times 1.00 ) \times \sin( ( s + ( \sin( s + s \times 0.99 ) \\
	&-4.27 ) \times -1.02 ) \times ( s + \sin( s + s -0.28 ) ) ) ) ) ) -1.08 ) \times ( \exp(s) \\
	&+ s ) ) ) \times ( \cos( \sin( ( \cos( \cos( \sin( ( s \times \sin( s ) + \sin( s ) + s + s ) \times s ) ) \\
	&-0.91 ) -1.07 ) \times ( s -4.76 ) ) + \cos( ( s + ( s -1.44 ) \times ( \cos( s \times s \\
	&+ \sin( s + 0.45 ) ) + 0.05 ) ) \times 1.59 \times s ) ) -1.00 ) )
\end{aligned}
$}
& {2}
& {112}
& {$3.046 \times 10^{-5}$}
\\
\hline
(d)
& PCSR without $\mathcal{L}_\text{mode}$
& {\small $
\begin{aligned}
	S_w &= ( \cos( \sin( \sin( s + \exp(( \cos( s \times ( s \times s + 0.97 + s ) \times s \times s ) + 1.75 ) \\
	&\times \sin( \exp(\exp(\sin( \exp(s) \times 0.89 ) + \sin( \sin( \exp(\exp(\sin( \sin( \sin( \sin( \sin( \\
	&\exp(\sin( \sin( \exp(\sin( \exp(\sin( s + 1.26 + ( \sin( \exp(s -0.42) \times \sin( s ) ) \\
	&+ \sin( s ) + s ) \times \sin( s \times 0.71 ) ) + \sin( ( \sin( \sin( \sin( s ) + \exp(s) ) ) + s ) \\
	&\times 2.27 ) \times 1.00) ) + 1.21) ) ) + \sin( \exp(s + s \times 1.26) )) \times 0.43 + s + 0.36 ) ) \\
	&+ 1.20 ) + \sin( ( s + \sin( \sin( s + \exp(\sin( s )) ) ) ) \times 2.40 ) \times 1.13 ) ))) \\
	&+ 1.37 ) ))) )) \times ( s -0.65 ) ) + 0.59 ) + 0.39 ) -0.18 ) \times 1.23 -0.001
\end{aligned}
$}
& {12}
& {134}
& {$3.321 \times 10^{-5}$}
\\
{}
& PCSR with $\mathcal{L}_\text{mode}$
& {\small $
\begin{aligned}
	S_w &= \sin( \exp(( s \times -0.88 + \exp(\sin( s \times s \times s \times 1.69 + s \times ( s + s ) \\
	&\times \exp(( s \times 0.47 + \exp(( s \times s \times -77.22 + 3.26 ) \times 6.57 + \sin( s ) \\
	&\times ( ( ( s + 6.57 ) \times 1.11 + \cos( s ) ) \times ( s \times s \times -77.22 + 3.26 ) + \cos( s \times s \\
	&\times -0.52 \times s ) + \sin( s + ( s + s ) \times -18.56 ) -17.47 ) \times 1.25) + s ) \times s \\
	&\times \sin( s ) \times s \times -1.00 + s \times \exp(( \sin( s ) \times ( s -4.27 ) -0.12 ) \times s \\
	&\times 1.77) \times s \times -17.47 \times ( s \times 4.05 + 0.87 )) \times \exp(s \times -1.77) \\
	&\times 26.08 -0.40 ) \times 7.98) \times -0.50 \times s ) \times s \times 3.41 + 0.37) ) \times 1.01
\end{aligned}
$}
& {2}
& {139}
& {$1.207 \times 10^{-4}$}
\\
\hline
\end{tabular}
\end{sidewaystable}

\begin{figure}[htbp]
\centering
\begin{subfigure}[b]{0.49\textwidth}
  \centering
  \includegraphics[width=\linewidth]{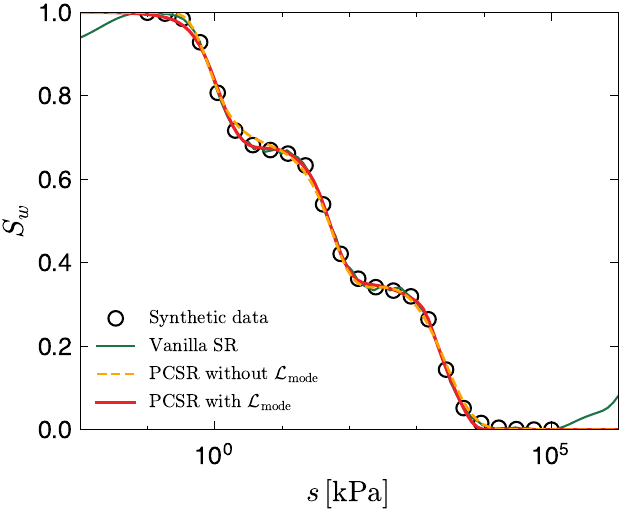}
  \caption{}
  \label{fig:multi_a}
\end{subfigure}
\hfill
\begin{subfigure}[b]{0.49\textwidth}
  \centering
  \includegraphics[width=\linewidth]{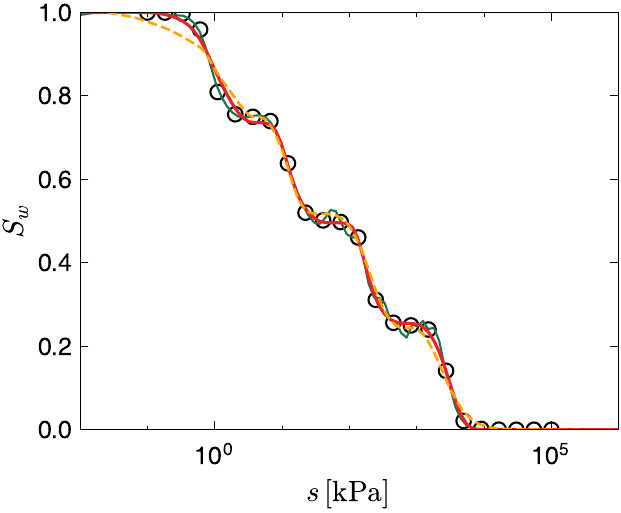}
  \caption{}
  \label{fig:multi_b}
\end{subfigure}
\caption{The symbolic model discovered from synthetically generated multimodal water retention data, compared against the \citep{durner1994hydraulic} model, vanilla SR, and PCSR without mode constraint.}
\label{fig:multimodal_wrc}

\end{figure}

Subsequently, Figure~\ref{fig:multimodal_wrc} illustrates the application of our PCSR framework to two additional water retention datasets with more complex curve shapes, specifically those with $N_\text{mode}=3$ and $4$. 
As experimental data for these multimodal water retention scenarios are not readily available, both datasets were synthetically generated using the \citep{durner1994hydraulic} model.

Although the synthetic datasets are designed to clearly exhibit the multimodal nature of the curves, the water retention curves discovered using vanilla SR result in noisy predictions that do not satisfy the physics constraints. 
In contrast, PCSR models, both with and without $\mathcal{L}_\text{mode}$, produce visually similar water retention curves. 
However, as shown in Figure~\ref{fig:multi_b}, the water retention curve without the mode constraint still fails to accurately capture the air entry pressure. 
Additionally, not incorporating the mode constraint leads to the discovery of water retention curves that do not match the desired number of modes, as shown in Table~\ref{tab:multi_fit_eqs}. 
In comparison, incorporating $\mathcal{L}_\text{mode}$ enables PCSR to reliably discover physically sound symbolic expressions that not only fit the given data points well but also adhere to the target number of modes.

The comparative analysis made in this section demonstrates that the PCSR framework, enhanced with the mode loss term, can effectively address the challenges of modeling multimodal water retention curves from sparse datasets. 
The water retention curves discovered using this approach consistently predict the correct number of modes and generate physically consistent symbolic expressions without requiring explicit parameter identification or predefined mathematical forms. 
These results underscore the effectiveness of the proposed PCSR framework for modeling water retention behavior of materials with complex, multi-scale pore structures. 

\begin{sidewaystable}
\caption{Discovered closed-form equations corresponding to those illustrated in Figures \ref{fig:multi_a}-\ref{fig:multi_b}.}
\label{tab:multi_fit_eqs}
\begin{tabular}{lllccc}
\hline
{}
& \textbf{Model}
& \textbf{Discovered expression}
& \textbf{$\hat{N}_\text{mode}$}
& \textbf{Complexity}
& \textbf{Loss}
\\
\hline
(a)
& PCSR without $\mathcal{L}_\text{mode}$
& {\small $
\begin{aligned}
	S_w &= \cos( ( \sin( ( \sin( s ) + ( \cos( \sin( ( \cos( s ) + s + 0.36 ) \times \sin( s ) ) + 0.38 ) \\
	&+ 0.07 ) \times \cos( ( \cos( \exp(s) ) + 1.10 ) \times ( \cos( s \times ( \cos( \cos( ( \cos( s \times 0.39 \\
	&+ ( s + 2.81 ) \times \cos( s + \sin( ( \cos( \cos( \sin( s ) ) + 2.02 ) + 3.12 ) \times s ) ) ) + 0.13 ) \\
	&\times ( ( \cos( \exp(\cos( s \times 1.94 )) ) \times \cos( s \times \cos( ( \cos( \cos( \cos( s ) ) + 2.02 ) \\
	&+ 3.12 ) \times ( s + 0.36 ) + 0.05 ) ) + 2.31 ) \times ( s \times \cos( \exp(s) ) + 1.00 ) \\
	&-0.36 ) ) ) + 2.33 ) \times ( \exp(\sin( \sin( s ) )) + 1.05 ) ) + s \times 0.89 + 0.98 ) ) ) \\
	&\times ( \cos( s ) + 1.40 ) ) + 0.69 ) \times \sin( s + s \times 0.49 ) + 0.04 )
\end{aligned}
$}
& {6}
& {127}
& {$1.026 \times 10^{-4}$}
\\[2ex]
{}
& PCSR with $\mathcal{L}_\text{mode}$
& {\small $
\begin{aligned}
	S_w &= \exp(\exp(( \sin( \exp(s) \times ( s -0.21 ) \times 2.31 ) + \sin( \sin( \sin( ( \exp(\cos( s ) \times s) \\
	&+ s ) \times \exp(\sin( \sin( s \times 0.79 ) )) + s \times 5.17 ) -4.25 ) + s \times 3.30 ) ) \times -2.64 \\
	&+ \sin( \sin( s ) )) \times s \times ( s \times ( \sin( s ) + \cos( \sin( ( s + \cos( \sin( ( ( \sin( s \times 3.29 \\
	&+ \exp(s \times 0.52) \times \exp(s + 1.00) + \sin( \sin( s \times 5.15 -2.70 ) ) ) + 0.96 ) \\
	&\times -2.63 + \sin( \sin( \sin( s ) ) ) + 1.55 ) \times s ) ) + 0.20 ) \times \sin( s \times 5.19 \\
	&+ ( ( s + \exp(\sin( \sin( s ) )) ) \times \exp(\sin( s \times ( s + 0.27 ) )) + s \times 3.30 ) \\
	&\times 1.02 -4.21 ) ) ) ) + s + s \times s ) \times -0.87) \times 1.00
\end{aligned}
$}
& {3}
& {136}
& {$3.861 \times 10^{-5}$}
\\
\hline
(b)
& PCSR without $\mathcal{L}_\text{mode}$
& {\small $
\begin{aligned}
	S_w &= ( \cos( \cos( \cos( s + s \times \cos( s ) \times ( ( \cos( \exp(s) \times s \times \exp(s \times s) \times 0.12 ) \\
	&-0.29 ) \times \cos( \cos( \cos( \exp(\cos( \cos( \sin( \cos( s \times \ln( \cos( \sin( s \times \exp(s \times 0.44 \\
	&+ \cos( s \times ( s \times 2.05 \times s \times -0.12 + 0.28 ) \times 3.90 )) ) + s ) ) ) ) ) + 1.00 \\
	&+ \exp(\exp(s \times 0.45 + \cos( \cos( s ) \times 4.13 \times ( s \times 1.29 -0.19 ) )) \times 0.94 \\
	&-0.31) )) ) \times -0.53 + \ln( \cos( \sin( s \times \exp(\cos( s \times ( s \times \exp(\cos( s )) \times s \\
	&\times -0.13 + 0.29 ) \times 3.91 ) + s \times 0.43) ) + s ) ) ) ) + \sin( s ) ) ) ) ) -0.54 ) \times 3.17
\end{aligned}
$}
& {8}
& {123}
& {$5.784 \times 10^{-4}$}
\\[2ex]
{}
& PCSR with $\mathcal{L}_\text{mode}$
& {\small $
\begin{aligned}
	S_w &= \exp(\exp(( s + \cos( \exp(s \times s \times 0.67 + 1.56) ) \times -0.38 ) \times 7.37) \\
	&\times \sin( \sin( s ) \times \sin( s \times \cos( \exp(\cos( ( \cos( \exp(\cos( \exp(s \\
	&\times \sin( \cos( \exp(( \cos( \cos( \cos( ( \exp(s \times 1.98) \times s \times 1.98 + s ) \times s ) ) \\
	&+ ( \cos( s ) + \cos( ( s + s ) \times -2.66 ) ) \times ( \exp(\exp(\cos( s \times s ))) + s ) ) + s \\
	&\times \cos( \exp(\cos( \cos( \exp(s \times \exp(s \times s + s) \times ( s -0.17 ) \times s) \\
	&\times -1.38 -1.10 ) )) ) \times 3.44 ) \times s + 0.56) \times ( s -1.10 ) ) ) \times s \\
	&+ \exp(s \times \sin( s ))) ) \times -0.58) ) + \cos( \cos( \exp(\cos( \cos( \exp(s) ) ) \\
	&+ \sin( s ) + 0.21) ) \times \exp(\exp(\sin( s ))) ) ) \times ( s -1.11 ) )) ) ) ))
\end{aligned}
$}
& {4}
& {139}
& {$1.361 \times 10^{-4}$}
\\
\hline
\end{tabular}
\end{sidewaystable}

\section{Conclusion and future outlook}
\label{sec:conc}
In this work, we introduce a physics-constrained symbolic regression (PCSR) framework designed to discover closed-form equations for water retention curves directly from sparse experimental data. 
By considering the data-driven discovery of physically consistent water retention curves as a multi-objective optimization problem, we formulate the symbolic regression task through an objective function comprising multiple terms derived from physics constraints. 
Assuming the number of modes in the pore size distribution of a material of interest is known a priori, our framework also incorporates a mode constraint to limit the number of convex-to-concave inflection points in the learned function, thereby effectively guiding the regression process. 
The PCSR framework has been rigorously evaluated on both unimodal and multimodal water retention datasets. 
Our results show that incorporating these additional constraints enables the discovery of mathematical expressions that not only fit the experimental data accurately but also describe water retention behavior in a thermodynamically consistent way, while preserving the desired number of modes. 
This advantage is particularly notable in cases where vanilla symbolic regression models tend to overfit and fail to reproduce the correct curve shape, especially for complex multimodal behaviors. 
Because the discovered models are expressed as closed-form equations, they can be directly integrated into existing hydraulic simulation workflows, just like conventional water retention models, offering the potential to improve simulations under heterogeneous and data-sparse conditions. 
While the present study focuses on meta-modeling of water retention in porous media rather than interpreting the physical significance of individual parameters or components within the discovered expressions, future research may explore embedding strategies to establish links between the symbolic structures and physical properties. 
Future work will also focus on extending PCSR to account for hysteretic water retention behavior, as well as exploring its application to broader geotechnical and environmental problems, and improving the efficiency of the symbolic regressor.

%
%
\section*{Open Research}
\setcurrentname{Open Research}\label{sec:data_avail}
The source code and dataset generated and/or analyzed during this study are publicly available at Zenodo \cite{yejin2026dataset}.

\section*{Acknowledgments}
\label{acknowledgements}
This work was supported by the National Research Foundation of Korea (NRF) grant funded by the Korea government (MSIT) (No. RS-2024-00360509), and the start-up grant from Case Western Reserve University.

\appendix
\renewcommand{\thesection}{\Alph{section}}
\renewcommand{\thesubsection}{\thesection.\arabic{subsection}}
\renewcommand{\thesubsubsection}{\thesubsection.\arabic{subsubsection}}
\renewcommand{\theequation}{\thesection.\arabic{equation}}
\renewcommand{\thefigure}{\thesection.\arabic{figure}}
\renewcommand{\thetable}{\thesection.\arabic{table}}

\setcounter{section}{0}
\setcounter{subsection}{0}
\setcounter{subsubsection}{0}
\setcounter{equation}{0}
\setcounter{figure}{0}
\setcounter{table}{0}

\titleformat{\section}
  {\large\bfseries}
  {Appendix~\thesection.}{1em}{}
  
\section{Effects of physics constraints on the discovered symbolic expressions}
\setcurrentname{Appendix A}\label{sec:apd}

This appendix examines how incorporating physics constraints influences the closed-form equation discovered by symbolic regression (SR). 
The three constraints considered in this work--monotonicity, limiting, and boundedness--are evaluated and compared against unconstrained (vanilla) SR under both noise-free and noisy settings. 
In each case, a dataset is generated from a benchmark equation specifically designed to highlight the effect of the corresponding constraint.

\subsection{Monotonicity constraint}
\label{app:A1}
\begin{figure}[htbp]
\centering
\begin{subfigure}[b]{0.49\textwidth}
  \centering
  \includegraphics[width=\linewidth]{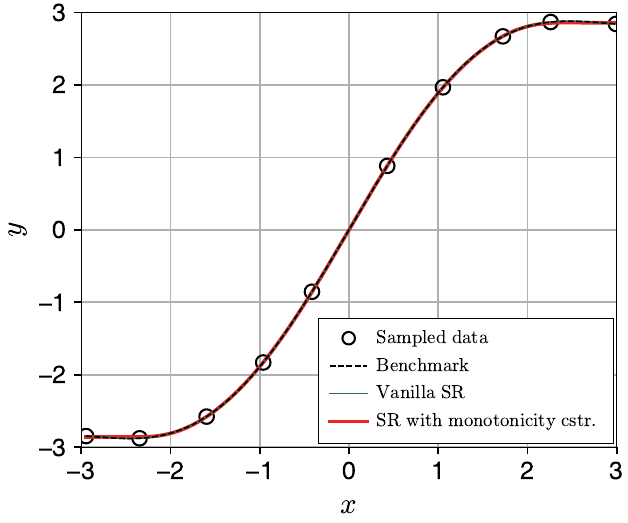}
  \caption{}
  \label{fig:mono_A1_a}
\end{subfigure}
\hfill
\begin{subfigure}[b]{0.49\textwidth}
  \centering
  \includegraphics[width=\linewidth]{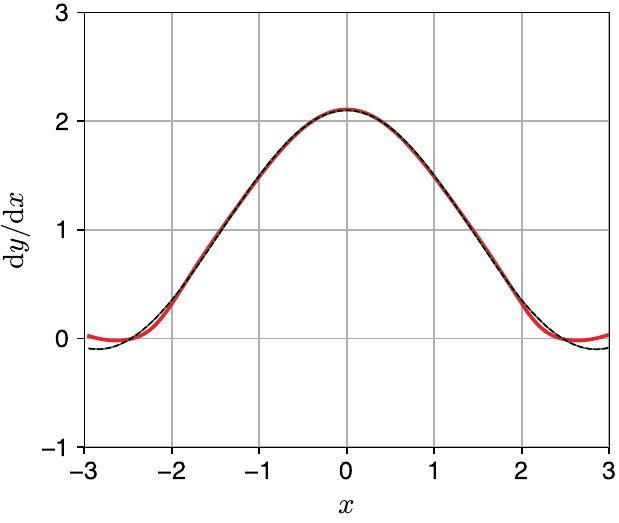}
  \caption{}
  \label{fig:mono_A1_b}
\end{subfigure}
\caption{Effect of the monotonicity constraint (${\mathrm{d} y}/{\mathrm{d} x} \ge 0$) in the noise-free case. (a) Benchmark equation and sampled data (black hollow circles) compared with vanilla SR (green solid curves) and SR with monotonicity constraint (red solid curves). (b) First derivatives of the corresponding equations, illustrating monotonicity violations in vanilla SR and preservation in the constrained model.}
\label{fig:mono_A1}
\end{figure}

To examine the effect of the monotonicity constraint, this example considers the benchmark equation:
\begin{equation}
\label{eq:eq_mono_A1}
y(x) = x + \sin(Ax) + \epsilon,
\end{equation}
which satisfies ${\mathrm{d} y}/{\mathrm{d} x} \ge 0$ for $A = 1$, but contains regions with ${\mathrm{d} y}/{\mathrm{d} x} < 0$ when $A > 1$. 
Here, $\epsilon$ denotes an additive noise term representing measurement or process uncertainty. 
By selecting $A = 1.1$ and $\epsilon = 0$, we generated uniformly sampled data (Figure~\ref{fig:mono_A1_a}). 
Both vanilla SR and SR with monotonicity constraint ${\mathrm{d} y}/{\mathrm{d} x} \ge 0$ were then applied to identify the mathematical expression that best fits the data.

To assess robustness in system structure identification, the search space was intentionally restricted: binary operators were limited to $\lbrace +, \times \rbrace$ and the unary operator to $\lbrace \exp \rbrace$, thereby excluding the trigonometric functions present in the benchmark equation. 
As summarized in Table~\ref{tab:A1_eqs}, both vanilla SR and SR with monotonicity constraint compensate for these restrictions by producing complex expressions composed of nested exponential terms.

Although both vanilla SR and SR with the monotonicity constraint produce curves that visually approximate the benchmark function, their derivative behaviors differ substantially. 
As shown in Figure~\ref{fig:mono_A1_b}, the vanilla SR model exhibits regions of negative slope, violating monotonicity. 
In contrast, the constrained SR model preserves nonnegative derivatives across the entire domain. 
Importantly, despite differences in symbolic representation, the constrained model correctly captures the intended behavior. 
This indicates that variations in the discovered expressions arise from the flexibility of symbolic regression under operator restrictions, rather than from instability or unreliability in the learned models. 

\begin{table}[htbp]
\caption{Discovered equations that best fit the data generated from the benchmark expression in Eq.~\eqref{eq:eq_mono_A1}, corresponding to those illustrated in Figure~\ref{fig:mono_A1}.}
\label{tab:A1_eqs}
\begin{tabular}{ll}
\hline
\textbf{Model} & \textbf{Discovered expression}
\\
\hline
Vanilla SR
& {\small
$\begin{aligned}
y 
&= 2 \times x \times ( 1.5181 \times \exp( x \times ( -0.0948 \times x + 6.829 \times 10^{-6} ) ) \\
& + ( 0.0327 \times x \times x - 9.953 \times 10^{-6} \times x - 0.4680 ) ) + 1.5125 \times 10^{-6}
	\end{aligned}$
}
\\
SR with monotonicity cstr.
& {\small
$\begin{aligned}
y 
&= ((\exp(((\exp(((x \times (-0.176)) \times (x + (\exp((x \times x) \times 0.469) \\
& \times (x \times 0.434)))) + 0.465) + ((x + x) \times x)) \\
& \times (-0.108)) + 0.502) + 0.718) \times x) - 0.001
	\end{aligned}$
}
\\
\hline
\end{tabular}
\end{table}

\begin{figure}[htbp]
\centering
\begin{subfigure}[b]{0.49\textwidth}
  \centering
  \includegraphics[width=\linewidth]{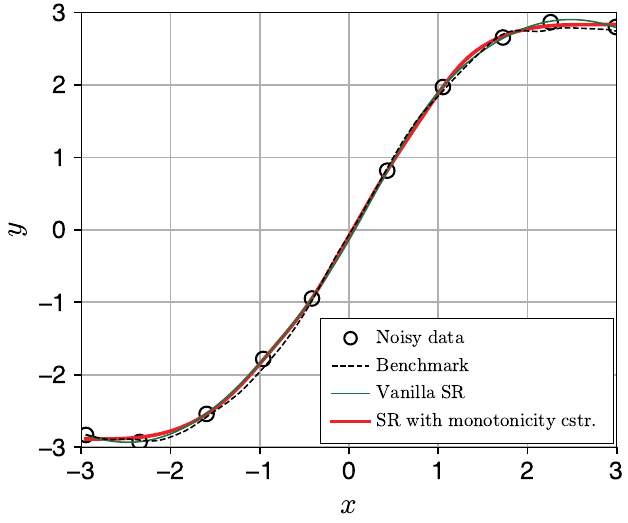}
  \caption{}
  \label{fig:mono_A2_a}
\end{subfigure}
\hfill
\begin{subfigure}[b]{0.49\textwidth}
  \centering
  \includegraphics[width=\linewidth]{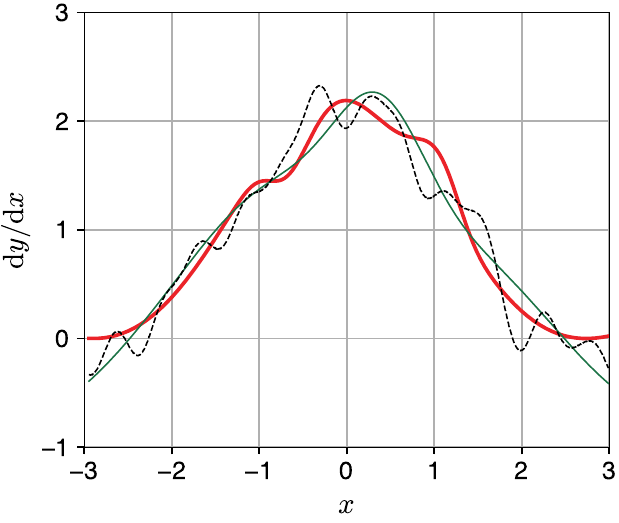}
  \caption{}
  \label{fig:mono_A2_b}
\end{subfigure}
\caption{Effect of the monotonicity constraint (${\mathrm{d} y}/{\mathrm{d} x} \ge 0$) in the presence of Gaussian process noise. (a) Noisy sampled data and learned equations from vanilla SR and SR with monotonicity constraint. (b) First derivatives of the corresponding equations, illustrating monotonicity violations in vanilla SR and preservation in the constrained model.}
\label{fig:mono_A2}
\end{figure}

When Gaussian process noise is introduced ($\epsilon = 0.05$; Figure~\ref{fig:mono_A2}), vanilla SR adapts more aggressively to the noisy data, resulting in expanded regions with negative derivatives. 
In contrast, SR with the monotonicity constraint strictly preserves monotonic behavior, demonstrating effective suppression of noise-induced violations. 

\subsection{Limiting constraint}
\label{app:A2}
To investigate the effect of limiting constraints, we trained both vanilla SR and SR with limiting constraint on a dataset sampled from the following benchmark:
\begin{equation}
\label{eq:eq_lim_A2}
y(x) = B(x + C) + \cos \left( (D(x + C) \right) + \epsilon. 
\end{equation}
This example considers the limiting constraint $y = 0$ for $x < 0.2$, while choosing $B = 0.4, C = 0.4, D = 3.0$, and $\epsilon = 0$ so that the sampled data from Eq.~\eqref{eq:eq_lim_A2} intentionally violate the imposed condition.

\begin{figure}[htbp]
\centering

\begin{subfigure}[b]{0.49\textwidth}
  \centering
  \includegraphics[width=\linewidth]{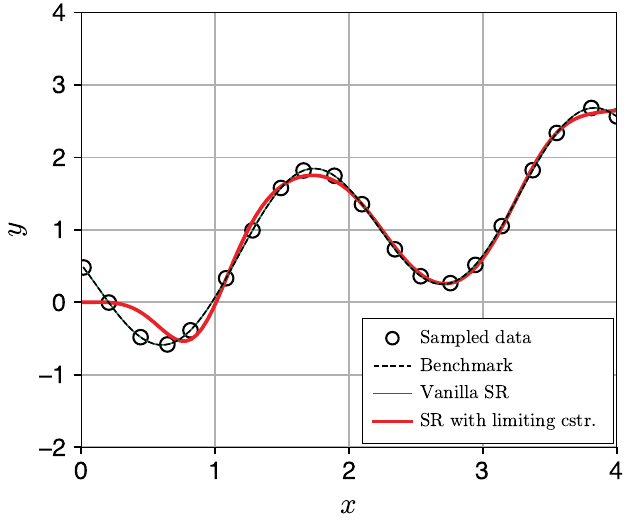}
  \caption{}
  \label{fig:lim_A3_a}
\end{subfigure}
\hfill
\begin{subfigure}[b]{0.49\textwidth}
  \centering
  \includegraphics[width=\linewidth]{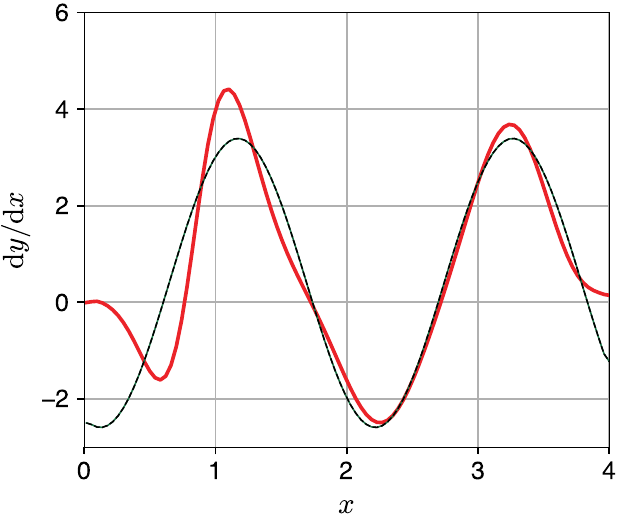}
  \caption{}
  \label{fig:lim_A3_b}
\end{subfigure}

\caption{Effect of the limiting constraint ($y = 0$ for $x < 0.2$) in the noise-free case. (a) Benchmark equation and sampled data compared with vanilla SR and SR with limiting constraint. (b) First derivatives of the corresponding equations, illustrating strict enforcement of the limiting behavior in the constrained model.}
\label{fig:lim_A3}

\end{figure}

As illustrated in Figure ~\ref{fig:lim_A3}, vanilla SR accurately fits the sampled data but produces nonzero values for $x < 0.2$, violating the limiting constraint. 
In contrast, SR with limiting constraint strictly enforces this condition while maintaining a good fit to the data elsewhere. 

\begin{figure}[htbp]
\centering
\begin{subfigure}[b]{0.49\textwidth}
  \centering
  \includegraphics[width=\linewidth]{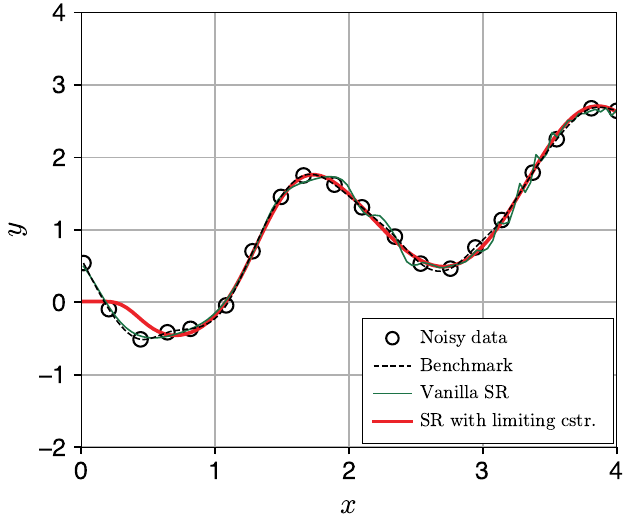}
  \caption{}
  \label{fig:lim_A4_a}
\end{subfigure}
\hfill
\begin{subfigure}[b]{0.49\textwidth}
  \centering
  \includegraphics[width=\linewidth]{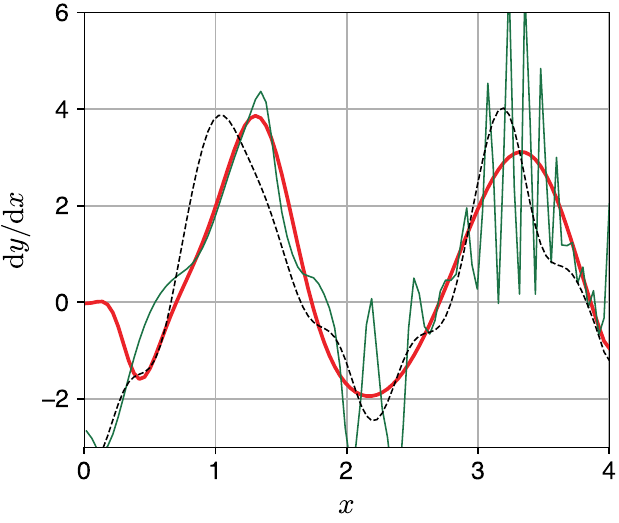}
  \caption{}
  \label{fig:lim_A4_b}
\end{subfigure}
\caption{Effect of the limiting constraint ($y = 0$ for $x < 0.2$) in the presence of Gaussian process noise. (a) Noisy sampled data and learned equations from vanilla SR and SR with limiting constraint. (b) First derivatives of the corresponding equations, demonstrating reduced sensitivity to noise under the imposed limiting constraint.}
\label{fig:lim_A4}
\end{figure}

For the case where the sampled data contain Gaussian process noise ($\epsilon = 0.2$), as shown in Figure~\ref{fig:lim_A4}, vanilla SR adapts strongly to noisy fluctuations, resulting in noticeable violations of the imposed limiting constraint. 
In contrast, SR with limiting constraint effectively suppresses overfitting in the limiting region, ensuring that the discovered expression remains consistent with the imposed condition. 

\subsection{Boundedness constraint}
\label{app:A3}
In this section, we investigate the effect of the boundedness constraint by considering the following benchmark equation:
\begin{equation}
\label{eq:eq_bound_A3}
y(x) = E (x^2 - F)(x^2 - G) + \epsilon.
\end{equation}
Based on this setting, we consider the boundedness constraint $0 \le y \le 1$, while deliberately choosing the parameters ($E = 0.125, F = 4.0, G = 2.25, \epsilon = 0$) such that the resulting benchmark equation--and thus the sampled data--violates this condition. 

\begin{figure}[htbp]
\centering
\begin{subfigure}[b]{0.49\textwidth}
  \centering
  \includegraphics[width=\linewidth]{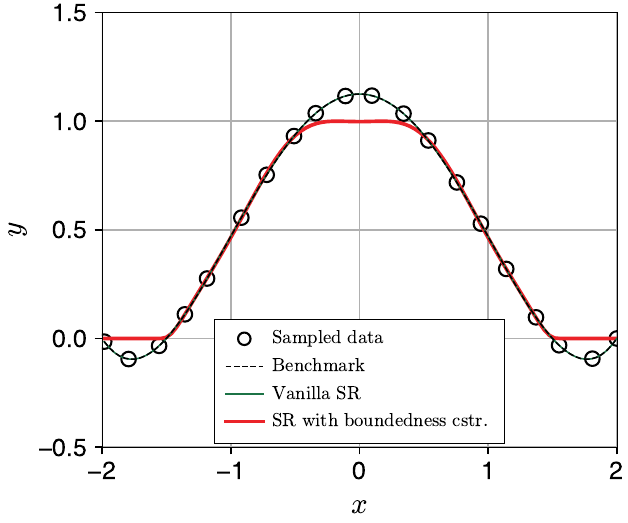}
  \caption{}
  \label{fig:bound_A5_a}
\end{subfigure}
\hfill
\begin{subfigure}[b]{0.49\textwidth}
  \centering
  \includegraphics[width=\linewidth]{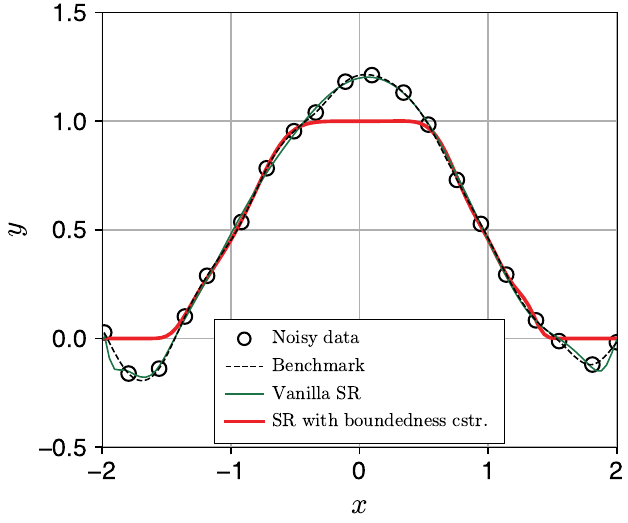}
  \caption{}
  \label{fig:bound_A5_b}
\end{subfigure}
\caption{Effect of the boundedness constraint ($0 \le y \le 1$) in (a) noise-free case and in (b) the presence of Gaussian process noise.}
\label{fig:bound_A5}
\end{figure}

As shown in Figure~\ref{fig:bound_A5_a}, vanilla SR discovers the expression that best fits the sampled data, while out-of-bound behavior consistent with the sampled data, while SR with boundedness constraint yields equation that ranges from $0 \le y \le 1$, satisfying the imposed boundedness constraint. 
A similar pattern is observed when Gaussian process noise is introduced to the dataset ($\epsilon = 0.5$). 
As illustrated in Figure~\ref{fig:bound_A5_b}, vanilla SR continues to reproduce out-of-bound deviations amplified by noise, whereas the constrained model consistently preserves the imposed bounds.

These results demonstrate that physically motivated constraints, particularly those derived from thermodynamic principles, can be effectively enforced within the symbolic regression framework. 
By embedding such constraints into the learning process, the proposed approach enables the data-driven discovery of models directly from data while preventing nonphysical behavior, even in the presence of noise.

\bibliographystyle{plainnat}
\bibliography{main}

\end{document}